\documentclass[aps,prb,twocolumn,superscriptaddress]{revtex4}
\usepackage{graphicx}% Include figure files
\usepackage{epsfig}
\usepackage{epstopdf}
\usepackage{subfigure}
\usepackage{tikz}
\usetikzlibrary{calc}

\usepackage{dcolumn}% Align table columns on decimal point
\usepackage{latexsym}
\usepackage{amssymb}
\usepackage{amsmath}
\usepackage{amsfonts}
\usepackage{wasysym}
\usepackage{bm}% bold math

\usepackage[colorlinks,bookmarks=false,citecolor=blue,linkcolor=red,urlcolor=blue]{hyperref}

\def\be{\begin{equation}}
\def\ee{\end{equation}}
\def\bea{\begin{eqnarray}}
\def\eea{\end{eqnarray}}

\def\vec{\mathbf}

\newcommand{\ua}{\uparrow}
\newcommand{\da}{\downarrow}
\usepackage{color}

\definecolor{darkblue}{rgb}{0,0.02,0.45}
\definecolor{darkred}{rgb}{0.45,0.02,0} 
\definecolor{darkgreen}{rgb}{0.02,0.45,0.0}

\usepackage{times}

\begin{document}
%\title{From quantum spin liquids to magnetically ordered phases in the Heisenberg model on the honeycomb lattice}
\title{Role of quantum fluctuations on spin liquids and ordered phases in the Heisenberg model on the honeycomb lattice}
%\title{The spin-$1/2$ and spin-$1$ $J_1-J_2-J_3$ Heisenberg model on the honeycomb lattice: a Schwinger boson mean field theory study}
\author{Jaime Merino}\email{jaime.merino@uam.es}
\affiliation{Condensed Matter Physics Center (IFIMAC) and Instituto Nicol\'as Cabrera, Universidad Aut\'onoma de Madrid, Madrid 28049, Spain}
\author{Arnaud Ralko}\email{arnaud.ralko@neel.cnrs.fr}
\affiliation{Institut N\'eel, UPR2940, Universit\'e Grenoble Alpes et CNRS, Grenoble, FR-38042 France}
\date{\today}

\begin{abstract}
% ----- motivation
Motivated by the rich physics of honeycomb magnetic materials, we obtain the
phase diagram and analyze magnetic properties of the spin-$1/2$ and spin-1
$J_1-J_2-J_3$ Heisenberg model on the honeycomb lattice.
% ----- approaches
Based on the SU(2) and SU(3) symmetry representations of the Schwinger boson
approach, which treats disordered spin liquids and magnetically ordered phases
on an equal footing, we obtain the complete phase diagrams in the $(J_2,J_3)$
plane. This is achieved using a fully unrestricted approach which does not
assume any pre-defined Ans\"atze.  
% ----- spin 1/2
For $S=1/2$, we find a quantum spin liquid (QSL) stabilized 
between the N\'eel, spiral and collinear antiferromagnetic phases in agreement with previous theoretical work. 
% ----- spin 1
However, by increasing $S$ from $1/2$ to $1$, the QSL is quickly destroyed due
to the weakening of quantum fluctuations indicating that the model already behaves
as a quasi-classical system. The dynamical structure factors and temperature
dependence of the magnetic susceptibility are obtained in order to characterize
all phases in the phase diagrams.  Moreover, motivated by the relevance of the
single-ion anisotropy, $D$, to various $S=1$ honeycomb compounds, we have
analyzed the destruction of magnetic order based on a SU(3)
representation of the Schwinger bosons.
%, [\onlinecite{Wang2017}]. 
%We 
%obtain the critical $D_c$ at which the transition from the N\'eel phase to
%the trivial quantum paramagnet occurs.
%
Our analysis provides a unified understanding of the magnetic
properties of honeycomb materials realizing the $J_1-J_2-J_3$ Heisenberg model from the strong quantum spin regime at $S=1/2$ to
the $S=1$ case. Neutron scattering and magnetic susceptibility 
experiments can be used to test the destruction of the QSL phase when replacing $S=1/2$ by 
$S=1$ localized moments in certain honeycomb compounds.\end{abstract}


\maketitle
\section{Introduction} 

Quantum magnetism on geometrically frustrated lattices is a very active field
of research due to the possibility of discovering new states of matter with
exotic properties\cite{balents}. In large spin systems which can be considered
as classical, frustration can lead to a large degeneracy of the ground state
manifold.  In sufficiently low spin systems, the quantum
mechanical zero point motion  can forbid long range magnetic order and produce
a quantum spin liquid state (QSL), a correlated state that
breaks no symmetry and possesses topological properties, possibly sustaining 
fractionalized excitations
\cite{Anderson1973,Fazekas1974,Anderson1987,Liang1988,Sachdev1992,Sandvik2005}.
Although the triangular lattice was first theoretically proposed by Anderson\cite{Anderson1973} as
an ideal benchmark to search for the QSL, it was soon found that the $S=1/2$
antiferromagnetic (AF) Heisenberg model on a triangular lattice is magnetically
ordered with a 120$^0$ arrangement of the spins.  However, longer range
exchange couplings and/or multiple exchange processes can destabilize the
magnetic order of the isotropic triangular model leading to a QSL\cite{merino1999,trumper1999,merino2014,holt2014}. 
%Other triangular-type lattices have been proposed as QSL candidates such as the
%Kagom\'e and honeycomb lattices.  
Despite the intense activity, only a small
number of triangular materials have been identified as possible candidates for QSL
behavior such as the layered organic materials\cite{kanoda2016}: $\kappa$-(BEDT-TTF)$_2$Cu$_2$(CN)$_3$, 
EtMe$_3$Sb[(Pd(dmit)$_2$]$_2$ and the Kagom\'e lattice\cite{norman2017} material 
Herbertsmithite, ZnCu$_3$(OH)$_5$Cl$_2$.
Hence, there is a need to find evidence for QSL behavior in more compounds. 
%The layered organic materials, $\kappa$-(BEDT-TTF)$_2$Cu$_2$(CN)$_3$
%and EtMe$_3$Sb[(Pd(dmit)$_2$]$_2$, donÕt display signs of magnetic order down
%to very low temperatures \cite{kanoda2016} in the magnetic susceptibility. The
%Kagom\'e lattice material Herbertsmithite, ZnCu$_3$(OH)$_5$Cl$_2$, is also
%non-ordered down to 1.6 K and the existence of fractional excitations (spinons)
%is indicated by the presence of a continuum in the spin excitation
%spectrum\cite{norman2017} probed by neutron scattering experiments.   

Honeycomb lattice materials have attracted lots of attention in recent years due to 
their interesting and poorly understood magnetic properties. Inorganic materials such as
Na$_2$Co$_2$TeO$_6$ \cite{lefrancoise2016},  BaM$_2$(XO$_4$)$_2$ (with X=As)
\cite{martin2012}, Bi$_3$Mn$_4$O$_{12}$(NO$_3$) \cite{smirnova2009} and
In$_3$Cu$_2$VO$_9$ \cite{yan2012} are examples of honeycomb lattice
antiferromagnets  in which the magnitude of the spin varies from $S=1/2$
in BaM$_2$(XO$_4$)$_2$  for M=Co to $S=1$ for M=Ni (with X=As) and
to $S=3/2$ in Bi$_3$Mn$_4$O$_{12}$(NO$_3$). The rather low magnetic ordering
temperature of $T_N=2$ K in the $S=1/2$ honeycomb antiferromagnet,
In$_3$Cu$_2$VO$_9$, and of only 5.35 K in BaCo$_2$(AsO$_4$)$_2$ suggest the
possible existence of QSLs in these compounds. Recent inelastic neutron scattering experiments 
indicate the presence of a QSL in $\alpha$-RuCl$_3$ \cite{banerjee2016,banerjee2017,do2017}, a material 
that realizes the Kitaev \cite{kitaev2005} quantum spin model on the honeycomb lattice.
Single-ion anisotropy of strength $D$ plays an important role in $S=1$ honeycomb
magnets such as: \cite{asai2016,asai2017} Ba$_2$NiTeO$_6$,  and in Mo$_3$S$_7$(dmit)$_3$ organometallic
compounds in which a relatively large $D$ can be induced by spin-orbit coupling.\cite{merino2016,khosla2017,merino2017,jacko2017}
%and where the Mo atoms form triangular clusters arranged in
%honeycomb layers with effective localized $S=1$ moments at each cluster
%site.\cite{merino2016,khosla2017,merino2017} 

%The interesting magnetic properties of honeycomb
%materials has attracted a lot of attention from theorists. 

It is important then to understand theoretically the 
magnetic properties of interacting localized moments on the frustrated honeycomb lattice as 
has been previously done on triangular lattices.  
Although the numerical evidence for a QSL in the half-filled Hubbard model on the honeycomb
lattice\cite{meng2010} has been questioned\cite{sorella2012}, exact
diagonalization studies on the $J_1-J_2$ Heisenberg model with $S=1/2$ have
found evidence for short range spin gapped phases for $J_2=0.3-0.35$ suggesting
the presence of a Resonance Valence Bond (RVB) state.\cite{lhuillier2001} The
possible existence of a magnetically disordered phase in this parameter regime
has been corroborated by more recent numerical work \cite{albuquerque2011,reuther2011},
including DMRG \cite{sheng2013} and series expansions. \cite{oitmaa2011}   Schwinger
boson mean-field theory (SBMFT)\cite{auerbach} is consistent with these predictions 
finding a magnetically disordered region between the N\'eel 
and spiral phases\cite{Lamas}. This disordered region consists of a
gapped QSL and a valence bond crystal (VBC) phase. In contrast to the $S=1/2$
model, the $S=1$, $J_1-J_2-J_3$ Heisenberg model on the honeycomb lattice
remains largely unexplored theoretically in spite of its relevance to several
materials as discussed above. DMRG studies suggest the existence of a spin disordered
region\cite{sheng2015} arising between the N\'eel and spiral phases even for
this larger $S=1$ case.

Motivated by recent successes of SBMFT in capturing important features of 
the spin-1/2, $J_1-J_2$ Heisenberg model on the honeycomb lattice we apply SBMFT
to the spin-1/2 and 1 $J_1-J_2-J_3$ Heisenberg model in the full $(J_2,J_3)$ 
parameter range. The SBMFT approach is particularly useful since it can describe ordered and
disordered phases on equal footing; the magnetically ordered phases resulting
from the condensation of the bosons at particular order wave vectors of the
system. We use the SU(2) formulation of the SBMFT in which the relevant
Heisenberg model with $S=1/2,1$ is expressed in terms of antiferromagnetic and
ferromagnetic bonds which are described through variational parameters
\cite{gazza1993,flint2009}.  We also introduce a SU(3)
formulation\cite{jaime2014} to describe the $S=1$ case which requires three
Schwinger bosons instead of the two of the SU(2) formulation.  The SU(3)
representation is used to adequately deal with the effect of single-ion anisotropy
in the Heisenberg model.  We have allowed for all possible point group
and translational symmetry breakings, keeping as many mean field parameters
which avoids biased guesses.  Our completely unrestricted solutions to the 
SBMFT equations has allowed us to obtain a consistent description of the
magnetic properties of $S=1/2$ and $S=1$ Heisenberg models in the full $(J_2,J_3)$ parameter range.
 
After obtaining the phase diagram  for both the $S=1/2$ and $S=1$ models, 
%we  QSL phase in the $S=1/2$ model 
%$J_3=0$ and around $J_3=J_2=0.5-0.6$.  
we find that a QSL phase is stable over a broad region of the $(J_2-J_3)$ phase diagram 
extending between the N\'eel, spiral phase and collinear antiferromagnet (CAF) phases consistent with previous numerical
work.  Our results agree with previous SBMFT studies restricted to $J_3=0$\cite{lamas2013} and to  
the $J_3=J_2$ line\cite{cabra2011}. Within SBMFT we find that the QSL region disappears in the $S=1$
model where a direct transition from the N\'eel to the spiral phase occurs. This indicates the 
fragility of the QSL phase as quantum fluctuations are reduced from $S=1/2$ to $S=1$, the latter 
behaving as cuasi-classical system.
We characterize the different phases by computing the dynamical spin structure
factor and the magnetic susceptibility in the different phases. Having in mind the $S=1$ materials we explore the effect of 
the single-ion anisotropy on the N\'eel order. We find that the N\'eel is destabilized at a sufficiently large
$D>D_c$, where a transition to a trivial paramagnet consisting on the tensor
product of $S_z=0$ states occurs. The critical single-ion anisotropy strength, $D_c$, 
is found to be rapidly suppressed by frustration suggesting a possible route to induce
a quantum paramagnetic phase in $S=1$ compounds such as Mo$_3$S$_7$(dmit)$_3$.

In Sec. \ref{sec:model} we introduce the frustrated $J_1-J_2-J_3$ Heisenberg model on the honeycomb lattice
we have studied and the SBMFT approach in the SU(2) representation we have used to solve the model. In Sec. \ref{sec:phased} we obtain the SBMFT phase diagrams of 
the $S=1/2$ and $S=1$ models comparing them in detail. The temperature dependence of the
magnetic susceptibilty is obtained and discussed in Sec. \ref{sec:magnetsus} whereas in Sec. \ref{sec:dynam}
we analyze the dynamic structure factor. In Sec. \ref{sec:singleion} 
single-ion anisotropy effects in the $S=1$  model are analyzed using the SU(3) slave boson
representation. We end up with some conclusions in Sec. \ref{sec:conclu}. 

\section{Model and methods}
\label{sec:model}

\begin{figure}[ht]
			\begin{center}
			\includegraphics[width=0.2\textwidth,clip]{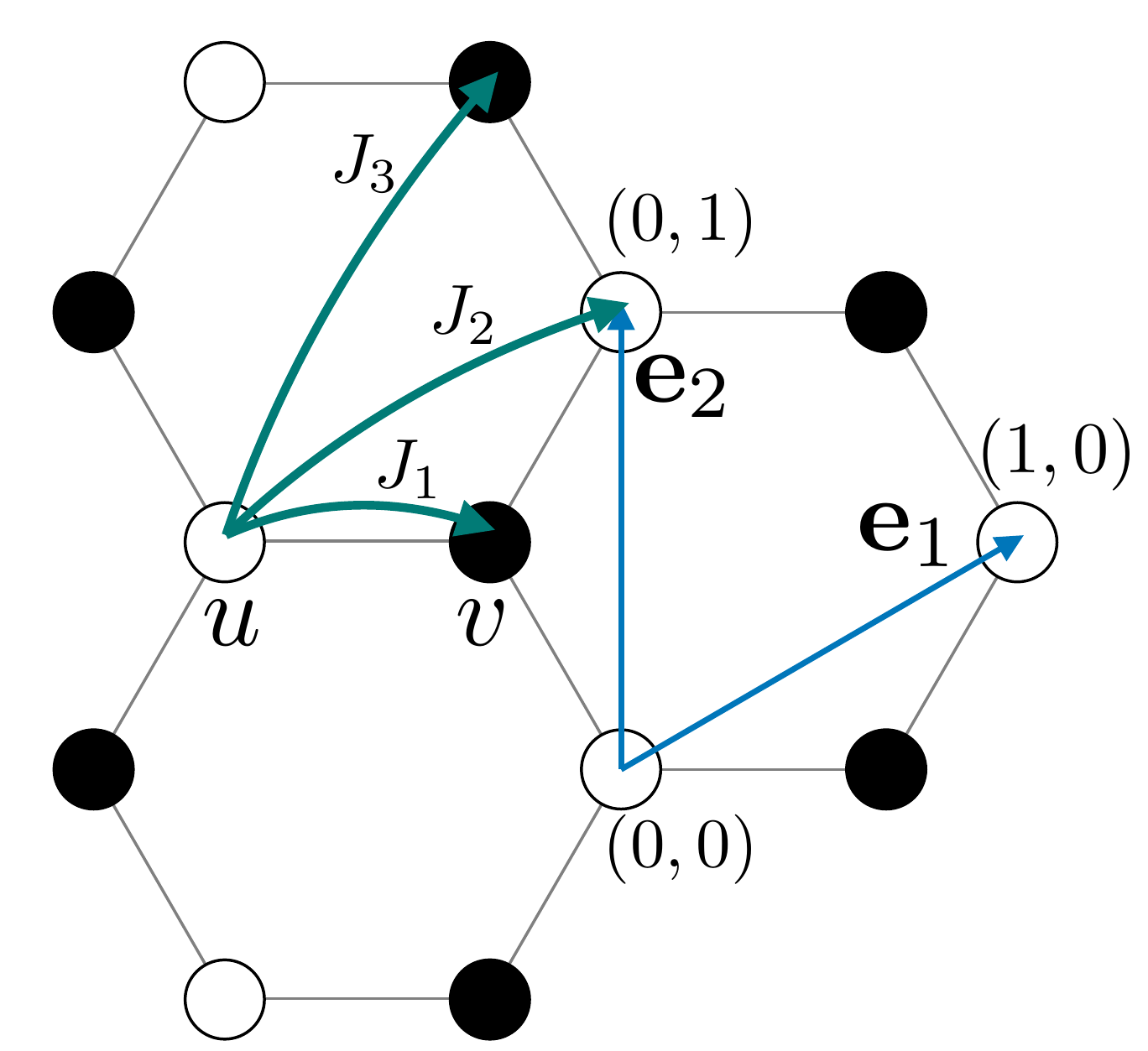}
			\caption{ The honeycomb lattice is defined by the translation
generator vectors $({\bf e}_1 , {\bf e}_2)$ and two sublattices $u$ and $v$.
The spin-spin exchange couplings up to third neighbors are depicted.
			\label{fig00}
			}
			\end{center}
			\end{figure}

The $J_1-J_2-J_3$ Heisenberg model is written:
\begin{eqnarray*}
{\cal H } &=& \sum_{i < j} J_{ij} S_{i} \cdot
S_{j}
\end{eqnarray*}
where $\vec{S}_{i}$ is the spin operator at site $i$, $J_{ij}$ the coupling
constant which is non zero only for first $(J_1)$, second  $(J_2)$ and third
$(J_3)$ neighbors, as summarized in Fig.~\ref{fig00},
together with the basic properties of the honeycomb lattice.
In order to study this Hamiltonian, we consider the Schwinger Boson mean field
theory that allows to treat on an equal footing disordered phases such like
quantum spin liquids (QSL) and magnetically ordered phases. The idea behind the
SBMFT is to express the spin operators in terms of bosons that carry the spin.
Usually, a SU(2) representation is considered, namely two bosonic flavors are
introduced to describe the spins.
%This is particularly well suited for spins 1/2 since two possible values can
%be taken along the quantization axis. 
%
This SU(2) representation is not restricted to spins 1/2 though, and the value
of the spin $S$ is controlled by a boson constraint that ensures the
commutation rule to be preserved under the transformation.
Following [\onlinecite{auerbach,Wang2010,Halimeh2016}], we introduce bosons that mimic the behavior of the spin
through the mapping:
\begin{eqnarray}
\vec{S}_{i} = \frac{1}{2} b_{i, \alpha}^{+} \vec{\tau}_{\alpha,\beta} b_{i, \beta}^{}
\end{eqnarray}
where $\vec{ \tau}$ are the Pauli matrices, $b_{i \sigma}^{+}$ the boson
creation operator of spin $\sigma$ on site $i$. As said, in order to preserve
the SU(2) commutation rule, the following local constraint has to be fulfilled
on each site:
\begin{eqnarray}
 b_{i \uparrow}^{+} b_{i \uparrow}^{} + b_{i\downarrow}^{+} b_{i \downarrow}^{} &=& 2 S.
\label{eq:bosonconstraint}
\end{eqnarray}
However, it is technically very hard to verify this
constraint exactly, thus it will be imposed on average on each site of the
lattice by introducing Lagrange multipliers $\mu_w$, with $w=u,v$ the
sublattice index.
			\begin{figure}[ht]
			\begin{center}
			\includegraphics[width=0.4\textwidth,clip]{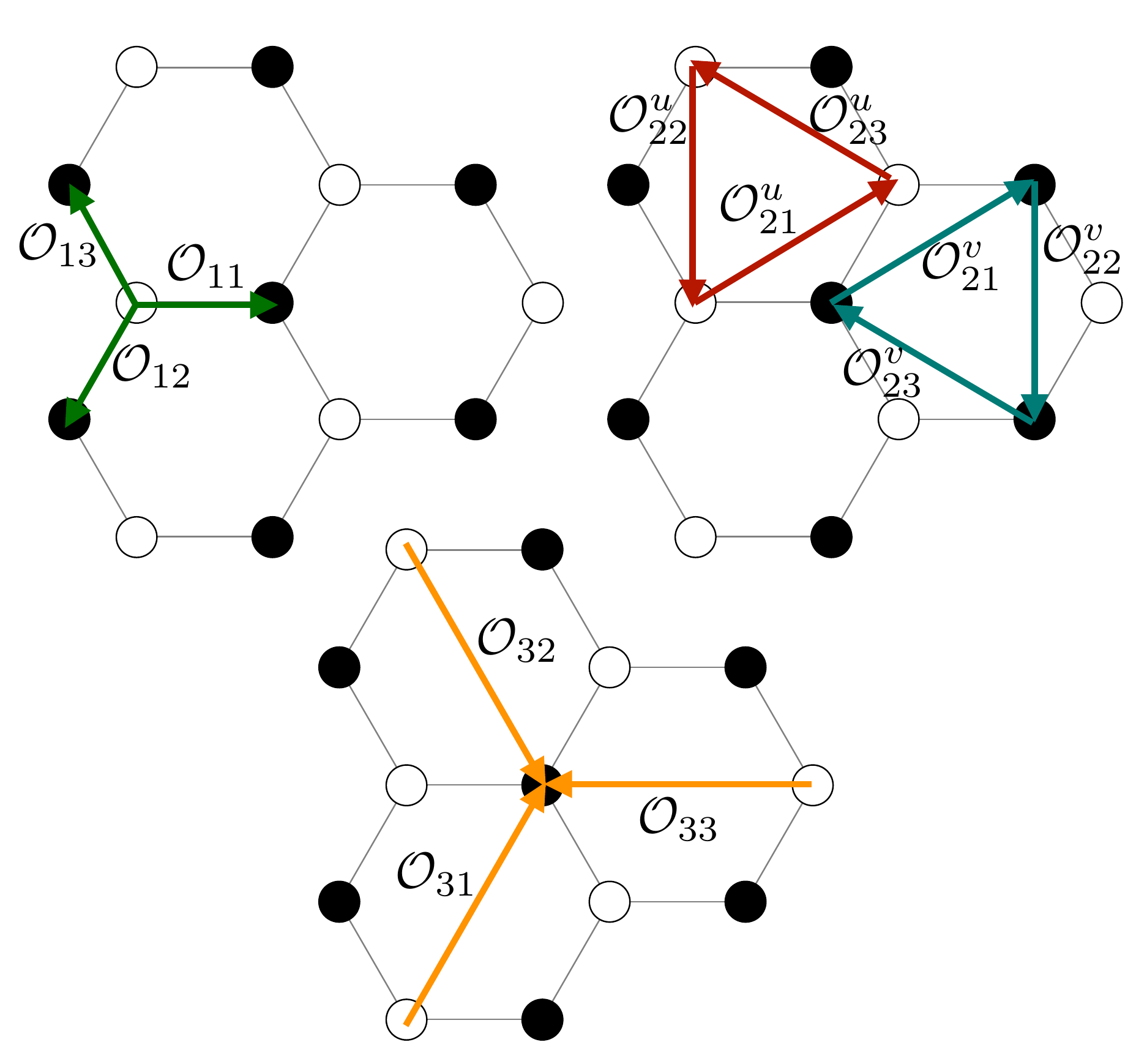}
            \caption{
				The twelve independent mean field complex parameters ${\cal
O}_{id}$ and their clockwise orientation conventions allowing point group
symmetry breaking on the lattice. The first subscript $i$ refers to the
neighbors (first, second and third), while the second $d$ refers to the three
directions. Note that for the second neighbors $i=2$, connected sites are on
the same sublattices, we then introduce two sets of mean field parameters
labelled with the extra subscript ${\cal O}^w$.
			\label{fig01}
			}
			\end{center}
			\end{figure}
One can introduce two SU(2) invariant quantities from which the Hamiltonian
could be rewritten \cite{flint2009}:
\begin{eqnarray}
	\hat{A}_{ij} &=& \frac{1}{2} \left[ b_{i \uparrow}^{ } b_{j \downarrow}^{} - b_{i \downarrow}^{} b_{j \uparrow}^{} \right], \\
	\hat{B}_{ij} &=& \frac{1}{2} \left[ b_{i \uparrow}^{+} b_{j \uparrow}^{} + b_{i \downarrow}^{+} b_{j \downarrow}^{} \right].
\end{eqnarray}
$ \hat{A}_{ij}^+$ creates a singlet on the oriented bond $(i,j)$ while $
\hat{B}_{ij}$ allows for spinon hopping on the same bond. It is clear from
these two quantities that the first one is favored in a gapped disordered phase
where spins are paired together as singlets, while the second needs an ordered
background to allow the spinon for hopping. It can be easily verified that:
\begin{eqnarray}
	\vec{S}_{i} \cdot \vec{S}_{j} &=& : \hat{B}_{ij}^+ \hat{B}_{ij} : - \hat{A}_{ij}^+\hat{A}_{ij } \nonumber \\
\end{eqnarray}
where $:\hat{O}:$ refers to the normal ordering. This allows for a simple
re-expression of the Hamiltonian, and after a mean field decoupling on
$\hat{A}$ and $\hat{B}$ operators:
\begin{eqnarray}
	 \hat{A}_{ij}^+ \hat{A}_{ij}  &\to& A_{ij}^* \hat{A}_{ij}  + \hat{A}_{ij}^+ A_{ij} - A_{ij}^* A_{ij} \\
	 \hat{B}_{ij}^+ \hat{B}_{ij}  &\to& B_{ij}^* \hat{B}_{ij}  + \hat{B}_{ij}^+ B_{ij} - B_{ij}^* B_{ij}
\end{eqnarray}
the final effective Hamiltonian is only expressed as bilinears of boson
operators. In this expression, $A$ and $B$ are complex mean field parameters to
be calculated from:
\begin{eqnarray}
	 A_{ij} &=& \langle \textrm{gs} | \hat{A}_{ij} | \textrm{gs} \rangle ~~,~~ 
	 B_{ij} = \langle \textrm{gs} | \hat{B}_{ij} | \textrm{gs} \rangle.
\label{eq:mfp}
\end{eqnarray}
Note that the ground state wave function $| \textrm{gs}
\rangle$ is the vacuum of the boson spectrum in this language, namely a state
without any Bose condensation.  On finite system a finite gap scaling as $\sim
1/\sqrt{n_c}$ is always present. This ensures us that the above definitions of
$A$ and $B$ are always verified.

Since only exchange couplings up to third neighbors are considered, that we
want to preserve the translational invariance of the solutions by allowing for
rotational symmetry breaking, we are ending up with 24 inequivalent mean field parameters called ${\cal O}_{id}$, 12 for $A$ and 12
for $B$. The first subscript $i$ refers the neighbor (1, 2 or 3) while the second $d$ to one of the three possible neighbors at distance $i$. This is summarized in Fig.\ref{fig01}, as well as the bond orientation we have used.

The final SU(2) Hamiltonian, up to a constant, is expressed as:
\begin{eqnarray}
	H &=& \sum_{ i < j } J_{ij} \left[ B_{ij}^* \hat{B}_{ij} + B_{ij} \hat{B}_{ij}^+ - A_{ij}^* \hat{A}_{ij} - A_{ij} \hat{A}_{ij}^+ \right] \nonumber \\
	&+& \sum_{i} \mu_i \left[ \sum_\sigma b_{i\sigma}^+ b_{i\sigma} - 2 S \right] - \langle H \rangle.
	\label{eq:mfh}
\end{eqnarray}

This mean field Hamiltonian can be block diagonalized by rewriting it in the
Fourier space. The unit cell of Fig.\ref{fig00}
contains two sites $w=u,v$ and any site $i$ of the lattice can be repaired by
the unit cell coordinate ${\bf r}$ and the sub lattice $w$. We then define
the Fourier transform of the boson operators as:
\begin{eqnarray}
	b_{{\bf{r}},w} &=& \frac{1}{\sqrt{n_c} } \sum_{ \bf q } e^{i {\bf q} \cdot {\bf r} } b_{ {\bf q } , w }
\end{eqnarray}
with $n_c$ the number of unit cells in the lattice. The mean field Hamiltonian then becomes:
\begin{eqnarray}
	H &=&   \sum_{\bf q }  \Psi_{\bf q}^+ M_{\bf q} \Psi_{\bf q} - (2S + 1 ) n_c \sum_w \mu_w - \langle H \rangle
\end{eqnarray}
with 
$\Psi_{\bf q} = ( b_{{\bf q} \uparrow}^{u} , b_{{\bf q} \uparrow}^{v} ,  b_{-{\bf q}
\downarrow}^{u+} , b_{-{\bf q} \downarrow}^{v+} )^T$ a four component Nambu
spinor, the 4$\times$4 matrix $M_q$ given by
\begin{widetext}
\begin{eqnarray*}
	M_{\bf q}&=&  \frac{1}{2} 
	\begin{bmatrix}
		J_2 ( B_{2d}^{u*} \phi_{2d} + B_{2d}^u {\phi}^*_{2d} ) + \mu_u & J_1 B_{1d} {\phi}^*_{1d} + J_3 B_{3d} {\phi}^*_{3d} & J_2 A_{2d}^{u*} ( {\phi}_{2d} -  {\phi}^*_{2d} ) & -J_1  A_{1d}^{*} {\phi}^*_{1d} -J_3  A_{3d}^{*} {\phi}^*_{3d}  \\
		J_1 B_{1d}^{*} \phi_{1d} + J_3 B_{3d}^{*} \phi_{3d}  & J_2 ( B_{2d}^{v*} \phi_{2d} + B_{2d}^v {\phi}^*_{2d} ) +\mu_v & J_1 A_{1d}^{*} \phi_{1d} + J_3 A_{3d}^{*} \phi_{3d} & J_2 A_{2d}^{v*} ( {\phi}_{2d} - {\phi}^*_{2d} ) \\
		J_2 A_{2d}^u ( {\phi}^{*}_{2d} - {\phi}_{2d}) & J_1 A_{1d} {\phi}^*_{1d} + J_3 A_{3d} {\phi}^*_{3d} & J_2 ( B_{2d}^{u*} \phi_{2d}^* + B_{2d}^u {\phi}_{2d} ) + \mu_u & J_1 B_{1d}^{*} {\phi}^*_{1d} + J_3 B_{3d}^{*} {\phi}^*_{3d}  \\
		- J_1 A_{1d} \phi_{1d} - J_3 A_{3d} \phi_{3d}  & J_2 A_{2d}^{v} ( {\phi}^*_{2d} - {\phi}_{2d} ) & J_1 B_{1d} {\phi}_{1d} + J_3 B_{3d} {\phi}_{3d} & J_2 ( B_{2d}^{v*} \phi_{2d}^* + B_{2d}^v {\phi}_{2d} ) + \mu_v
	\end{bmatrix}
\end{eqnarray*}
\end{widetext}

where a summation over repeated indices is assumed, and $\phi_{i,d}({\bf q}) =
e^{i {\bf q} \cdot {\bf \delta}_{i,d}}$ the phase factor induced between two
neighboring sites of distance $\delta_{i,d}$ from i$^\text{th}$ neighbors (1, 2
or 3) and in one of the three directions $d$ as displayed in Fig.\ref{fig01}.

Now, we perform a Bogolioubov transformation of the matrix $M_{\bf q}$ which
preserves the bosonic communication relations \cite{Colpa,Halimeh2016}, by
defining the new bosonic operators $\Gamma_{\bf q} = ( \gamma_{{\bf q} \uparrow}^{u}
, \gamma_{{\bf q} \uparrow}^{v} ,  \gamma_{-{\bf q} \downarrow}^{u+} , \gamma_{-{\bf q}
\downarrow}^{v+} )^T$ in such a way that $\Psi_{\bf q} = T_{\bf q} \Gamma_{\bf q}$. The mean
field Hamiltonian then takes the diagonal form

\begin{eqnarray}
	H &=&   \sum_{\bf q }  \Gamma_{\bf q}^+ \omega_{\bf q} \Gamma_{\bf q} - (2S + 1 ) n_c \sum_w \mu_w - \langle H \rangle
\end{eqnarray}
and the matrix $T_{\bf q}$ verifies the following conditions:
\begin{eqnarray}
	T_{\bf q}^+ \tau^4 T_{\bf q} = \tau^4, \\
	T_{\bf q}^+ M_{\bf q} T_{\bf q} = \omega_{\bf q},
\end{eqnarray}
where 
\begin{eqnarray}
	\tau^4 &=& 
	\begin{bmatrix}
		{\cal I}_2 & \\
		& -{\cal I}_2
	\end{bmatrix}
	, ~~~~~ \omega_{\bf q} = 
	\begin{bmatrix}
		\epsilon_{+} & \\
		& \epsilon_{-}
	\end{bmatrix}
\end{eqnarray}
with ${\cal I}_2$ the identity matrix of dimension 2, and $\epsilon_{-} = -
\epsilon_{+}$ if time reversal symmetry is preserved \cite{Halimeh2016}.

The Bogolioubov transformation matrix $T_{\bf q}$ takes then the specific block form 
\begin{eqnarray}
	T_{\bf q} &=& 
	\begin{bmatrix}
		U_{\bf q} & X_{\bf q} \\
		V_{\bf q} & Y_{\bf q}
	\end{bmatrix}.
\end{eqnarray}
Note that an elegant way of finding the Bogolioubov matrix $T_{\bf q}$ is to
consider a Choleski decomposition as detailed in [\onlinecite{Toth}]. Now that
the mean field Hamiltonian is diagonalized for any ${\bf q}$ point, one can
search for a fixed point in the mean field parameter space by minimizing the
free energy
\begin{eqnarray}
	{\cal F}_{\text{MF}} &=& \sum_{\bf q} \sum_w \epsilon_{q,\uparrow}^w - (2S + 1 ) n_c \sum_w \mu_w - \langle H \rangle,
	\label{eqn:fe}
\end{eqnarray}
with respect to the mean field parameters and the chemical potentials:
\begin{eqnarray}
	\frac{ \partial {\cal F}_{\text{MF}}}{ \partial {\cal O}_{id}} = 0, ~~~~~ \frac{ \partial {\cal F}_{\text{MF}}}{ \partial \mu_w} = 0.
\end{eqnarray}
This gives rise to a set of self-consistent equations that are numerically
solved.  In the same spirit, it is also possible to solve the self consistency
by computing at each step the mean field parameters in the gapped ground state
as defined in Eq.~\ref{eq:mfp}. As pointed out, since we are working on finite
systems, an artificial gap is always present even if the ground state at the
thermodynamic limit is gapless. This can be used in order to simplify and
evaluate Eq.~\ref{eq:mfp}.

The advantage of this procedure instead of minimizing the free energy is that
no numerical derivative is required. Moreover, the complexness of the mean
field parameters is naturally taken into account, which can be of importance if
a flux phase is the ground state. Finally, it allows for finding completely
unrestricted solutions. However, it is worth emphasizing that using both
procedures, we have always obtained same solutions in the present phase
diagrams. 

The minimization procedure is as follow. First, we start from a given {\it
ans\"atz} for the mean field parameters $\{\cal O\}$.
Depending the nature of the ground state, this {\it ans\"atz} has to be
carefully chosen for helping to a good convergence of the self consistency.
This is particularly true in regions of non-commensurate phases as described
below.

Plugging the solutions obtained in the large $S$ limit, classical solutions of
the Hamiltonian described in the next section, helps us to always find good solutions in
any part of the phase diagram.

Then, starting from high value of the chemical potentials and decreasing it, we 
fulfill the boson constraint of Eq.~\ref{eq:bosonconstraint}. Once a set of $\{
\mu_w \}$ is obtained, we diagonalize the mean field Hamiltonian and compute the
new set of $\{ \cal O \}$ by employing one of the two approaches presented
above (derivative of the free energy or explicit computation of the mean field
parameters).  Then we reconstruct the new Hamiltonian and continue this
algorithm until convergence up to an arbitrary tolerance. In our case, the
tolerance on the energy is at least $\sim 10^{-12}$ and on the mean field
parameters at least $\sim 10^{-9}$.

\section{Phase diagram}
\label{sec:phased} 
 
We now obtain and analyze the phase diagrams of the $S=1/2$ and $S=1$ $J_1-J_2-J_3$ 
Heisenberg model on the frustrated honeycomb lattice. Before discussing the model 
using the SBMFT approach we briefly revisit the classical phase diagram. 

\subsection{Classical phase diagram}
The classical phase diagram of the $J_1-J_2-J_3$ antiferromagnetic Heisenberg
model on a honeycomb lattice has been discussed in the literature
\cite{lhuillier2001} and we recall here the main results. The classical spin
$S$, at unit cell ${\bf r}$ of sublattice $w$ is given by: 
 \begin{equation}
{\bf S}_{rw} \equiv S \cos({\bf Q} \cdot {\bf r}+\phi_w)   {\bf e}_1 + S \sin({\bf Q} \cdot {\bf r}+\phi_w) {\bf e}_2
\end{equation}
where
%$w=u,v$ denotes the two sublattices, ${\bf
%r}_i$ is a Bravais vector of the triangular lattice and 
${\bf Q}$ denotes the magnetic ordering pattern.

We take $\phi_v=\phi$  and $\phi_u=0$, so $\phi$ is 
the relative phase between the two sublattices $u$ and $v$.  The classical energy per unit cell
reads:
\begin{widetext}
\begin{eqnarray}
\frac{ E_{\text{class}} }{ S^2 n_c } &=& {J_1 \over 2} \left[ \cos(\phi)  +\cos(\phi-Q_1) +\cos((Q_1-Q_2)+\phi) \right]+ J_2 \left[ \cos(Q_1)+\cos(Q_2)+\cos(Q_2-Q_1) \right]
\nonumber \\
&+&{J_3 \over 2} \left[ \cos(\phi+Q_1) +\cos(\phi-Q_1)+\cos(\phi+(Q_1-2 Q_2)) \right].
\end{eqnarray}
\end{widetext}

The phase diagram, in the $({J_2},{J_3})$ plane (in unit of
$J_1$) consists\cite{rastelli1979,lhuillier2001} of a N\'eel ordered phase with
${\bf Q}=(0,0)$ ($\phi=\pi$), a collinear antiferromagnetic phase with
${\bf Q}=(0,\pi)$ ($\phi=\pi$)  and a spiral phase with: ${\bf Q}=(Q_1,Q_1/2)$
where $Q_1=2 \arccos({J_1/2-J_2 \over 2 J_2-2 J_3})$ ($\phi=\pi$).
The
transition lines separating these phases are: (i) $J_3/J_1={1 \over 4} (-1 + 6
{J_2 \over J_1})$ between the N\'eel and spiral phases.  (ii) $J_2=0.5$ between
the N\'eel and the CAF phase for $J_3/J_1>0.5$.  (iii) 
$J_3/J_1={1 \over 4} (1+ 2 {J_2 \over J_1})$ between N\'eel and spiral phases. 
They are 
displayed in Fig.~\ref{fig:fig0} as thin continuous lines to make comparison
with the present study.
For $J_3=0$, an infinitely degenerate collection of spiral states
arises\cite{lhuillier2001,mulder2010} in the parameter range $J_2/J_1  \in [
1/6, 1/2]$.
The corresponding magnetic ordering vector ${\bf Q}^*$,  satisfies:
\begin{eqnarray} \cos(Q^*_1) + \cos(Q^*_2) + \cos(Q^*_1-Q^*_2) = {1
\over 8 (J_2/J_1)^2} -{ 3 \over 2}, \nonumber \\ \label{eq:Q} \end{eqnarray}
with the phase given by the equation:  \begin{eqnarray} \tan(\phi)={\sin(Q^*_2)
+ \sin(Q^*_1+Q^*_2) \over 1 + \cos(Q^*_2) + \cos(Q^*_1+Q^*_2)}, 
\end{eqnarray}
and $Q^*_1$ and $Q^*_2$ obtained from Eq.~(\ref{eq:Q}).
Linear order quantum fluctuations are found to diverge for $J_2/J_1 \gtrsim
0.1$ signaling the destruction of N\'eel order with no spiral magnetic order.
The singular behavior of quantum fluctuations is due to the infinite degeneracy
of the planar states found in the classical solution.
Classically, only when, $J_2 \rightarrow \infty$, ($J_3=0$) the 120$^0$
magnetic order is stabilized since the honeycomb lattice decouples into two
isotropic triangular lattices in this limit. However, we show below how quantum fluctuations 
actually stabilize
the 120$^0$ order in a region of $(J_2,J_3)$ in which it is not expected
classically. Nevertheless, the 120$^0$ solution is actually part of the spiral
states with wave vectors at the corner of the Brillouin zone.  
   
 \subsection{Quantum fluctuation effects}  
\begin{figure}[h]
   \centering
    \includegraphics[width=6cm]{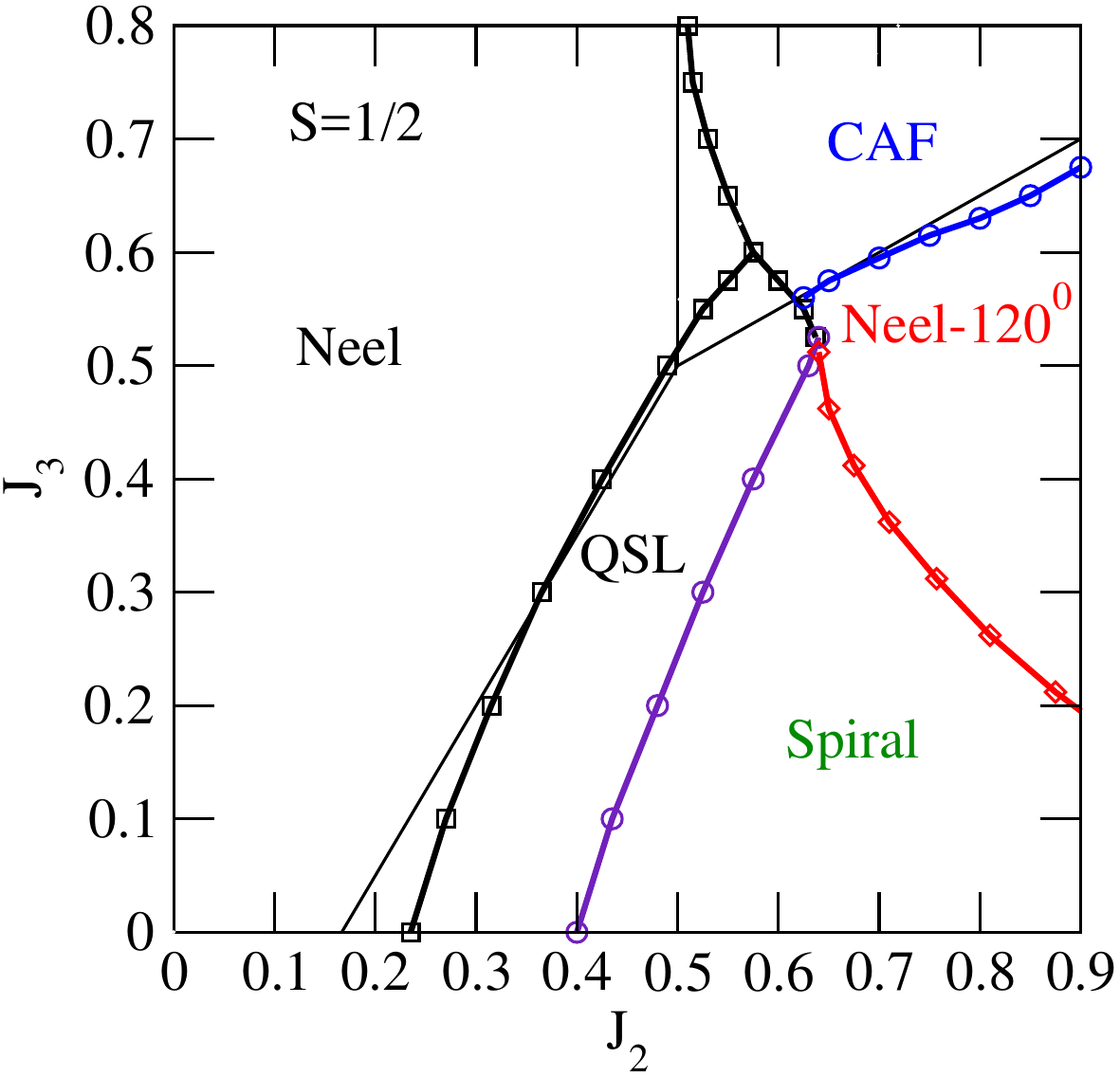}
    \includegraphics[width=6cm]{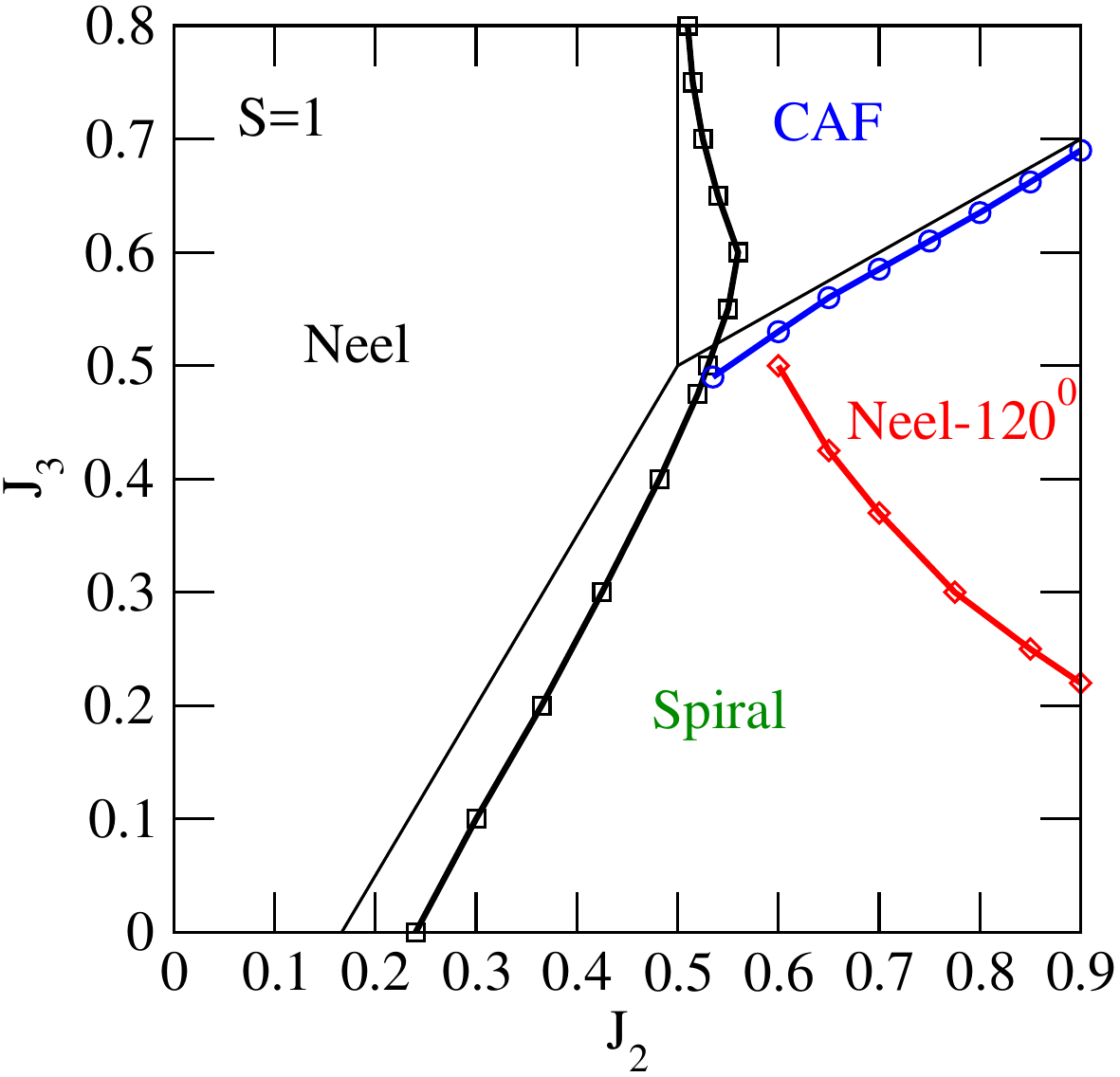}
	   \caption{Phase diagram of the $J_1-J_2-J_3$ Heisenberg model on the
honeycomb lattice. The $J_3$ vs $J_2$ phase diagram of the model obtained  from
SU(2)-SBMFT for $S=1/2$ is compared with the $S=1$ case.  There are three types
of magnetically ordered phases:  N\'eel antiferromagnet, spiral and collinear
antiferromagnet (CAF). The quantum spin liquid (QSL) previously found within
SBMFT for $J_3=0$ and around $J_3=J_2/2$ is found in a broad region close to
the boundary between the antiferromagnetic and spiral phases only for $S=1/2$.
For $S=1$ the QSL disappears and a direct transition from the N\'eel to the
spiral phase occurs. Within the spiral phase, the N\'eel-120$^0$ order is
favored by a sufficiently large $J_3$. The thin black lines show the classical
phase diagram for comparison. We have taken $J_1=1$ in the plot.}
 \label{fig:fig0}
 \end{figure}

We now discuss the effect of quantum fluctuations on the classical phase
diagram of the model using SU(2)-SBMFT.  The $S=1/2$ case has been considered
previously in the literature for $J_3=0$\cite{lamas2013} and $J_3=J_2$
\cite{cabra2011}. Here, we extend these studies to the whole $J_3$-$J_2$ plane
for both $S=1/2$ and $S=1$.
It is important to recall here that our mean field solutions are completely
unrestricted, in the sense that no particular symmetry is fixed a priori in
order to find the most general ones.

 In Fig. \ref{fig:fig0}, the SU(2)-SBMFT phase diagrams in the ($J_2$,$J_3$)
plane  for $S=1/2$ and $S=1$ are shown.
In both cases, there are regions of the
phase diagram in which three different classical configurations discussed above
are stabilized: the N\'eel, the spiral and the collinear antiferromagnet. These results
are obtainedon clusters up to 36x36 sites.

We recall that the classical transition lines between these phases are shown in
Fig. \ref{fig:fig0} as a guide.  As expected, the quantum fluctuations included
in the SBMFT also lead to magnetic ordering vectors, ${\bf Q}$, which are
different from the classical values and also select a particular configuration
from a classically degenerate manifold, the {\it order} from {\it disorder}
effect. For instance, we find the N\'eel-120$^0$  (or close to this order) in
a broad region of the phase diagram.  Although this spiral order wavevector
has the classical form: ${\bf Q}=(Q_1,Q_1/2)$, the magnitude, $Q_1$,
differs from the classical value in this $(J_2,J_3)$ region. {\it i. e.} $Q_1 \neq 2
\arccos({J_1/2-J_2 \over 2 J_2-2 J_3})$. This can be understood from the
quantum fluctuations leading to a magnetically ordered phase different from the
classical one as discussed previously.

{\it The $S=1/2$ case.} For $J_3=0$, quantum fluctuations are
found to stabilize the N\'eel phase above the classical critical ratio $J_2/J_1
\approx 0.2$, recovering previous findings.  \cite{lamas2013} Spiral order is
found at larger ratios, $J_2/J_1 > 0.4$, and a gapped quantum spin liquid
occurs between the N\'eel and spiral order for $0.21 < J_2/J_1 < 0.37$.
Quantum fluctuations select a particular wavevector ${\bf Q}$ from the infinite
classical manifold defined by Eq. \ref{eq:Q} as found previously using
linear spin wave theory \cite{mulder2010}.  Also a staggered valence bond
crystal (SVBC) which breaks the rotational $C_3$ symmetry of the lattice occurs
between the QSL and the spiral phases in a quite narrow parameter range: $0.37
< J_2/J_1 < 0.4$.  We have recovered this phase
\cite{lamas2013} but due to its very tiny extension, it is almost not visible
in our phase diagram and we chose not to display it. The QSL found with SBMFT
is qualitatively consistent with DMRG studies in which a non-magnetically
ordered phase occurs in the range: $0.22 < J_2/J_1 < 0.25$.  \cite{sheng2013}
Also a plaquette valence bond crystal\cite{white2013,sheng2013} (PVBC) is found
for $0.25 < J_2/J_1< 0.35$ between the N\'eel and a SVBC \cite{white2013}
which differs from the gapped Z$_2$ QSL predicted by SU(2)-SBMFT. For non-zero
$J_3$, the $S=1/2$ SU(2)-SBMFT of Fig. \ref{fig:fig0}  shows how the QSL for
$J_2=0$ is robust in a broad region located between the N\'eel, spiral, and CAF
phases. This is in good agreement with the QSL found by Cabra et. al.
\cite{cabra2011} but only along the $J_3=J_2$ line when $0.41 < J/2/J_1 < 0.6$.
Our SBMFT phase diagram is also in very good agreement with a
recent numerical analysis using exact diagonalization (ED) (see Fig. 2 of Ref.
[\onlinecite{albuquerque2011}] for instance), with the phase diagram
obtained by pseudo-fermion functional renormalization group (pf-FRG) approach
(see Fig. 1 of Ref. [\onlinecite{reuther2011}]) and with the coupled cluster method
(see Fig. 2 of Ref. [\onlinecite{bishop2012}]). Series
expansions\cite{oitmaa2011}  also find a magnetically disordered phase around
$J_3=J_2=0.5J_1$ whereas it is inconclusive for other $J_3/J_2$ ratios. 

{\it The $S=1$ case.} The SU(2)-SBMFT
$(J_2,J_3)$ phase diagram for $S=1$ is shown in
the lower panel of Fig. \ref{fig:fig0}.  The smaller effect of
quantum fluctuations compared to $S=1/2$ is evident from the shift
of the SBMFT lines towards the classical transition lines, as well as the
disappearance of the QSL phase. DMRG studies\cite{sheng2015} of the $S=1$ model
with $J_3=0$ do suggest the existence of a non magnetic disordered phase
(possibly a PVBC) in the parameter range $0.27< J_2/J_1 < 0.32$ between the
N\'eel and a stripe AF phase, and a 
magnetically disordered phase is found between $0.25 < J_2/J_1 < 0.34$ 
for $S=1$ using the coupled cluster method. \cite{bishop2016}
In our analysis, we just find a direct transition from the N\'eel to the spiral
phase with no intermediate magnetically disordered phase. It is worth
noticing that a careful analysis of the energy of the PVBC solution has been
performed, and we have always found that, in the SU(2)-SBMFT description, this
solution was slightly above either the N\'eel and the spiral solutions.
Finally, one can see the natural tendency of the boundary lines as $S$
increases to become closer and closer to the classical lines, at the exception
of the $120^o$ line as already mentioned. As the system is reaching the
classical limit, magnetic orders are strengthening until being ideal classical
solutions at very large $S$. Anticipating next sections, this
explains  why the branches observed in the dynamical structure factors for the
$S=1/2$ case are blurry in the quantum regime, due to quantum fluctuations,
where it should have been very sharp from a linear spin wave theory for
example.

\section{Magnetic susceptibility}
\label{sec:magnetsus}

\begin{figure}[ht]
   \centering
\includegraphics[width=4cm]{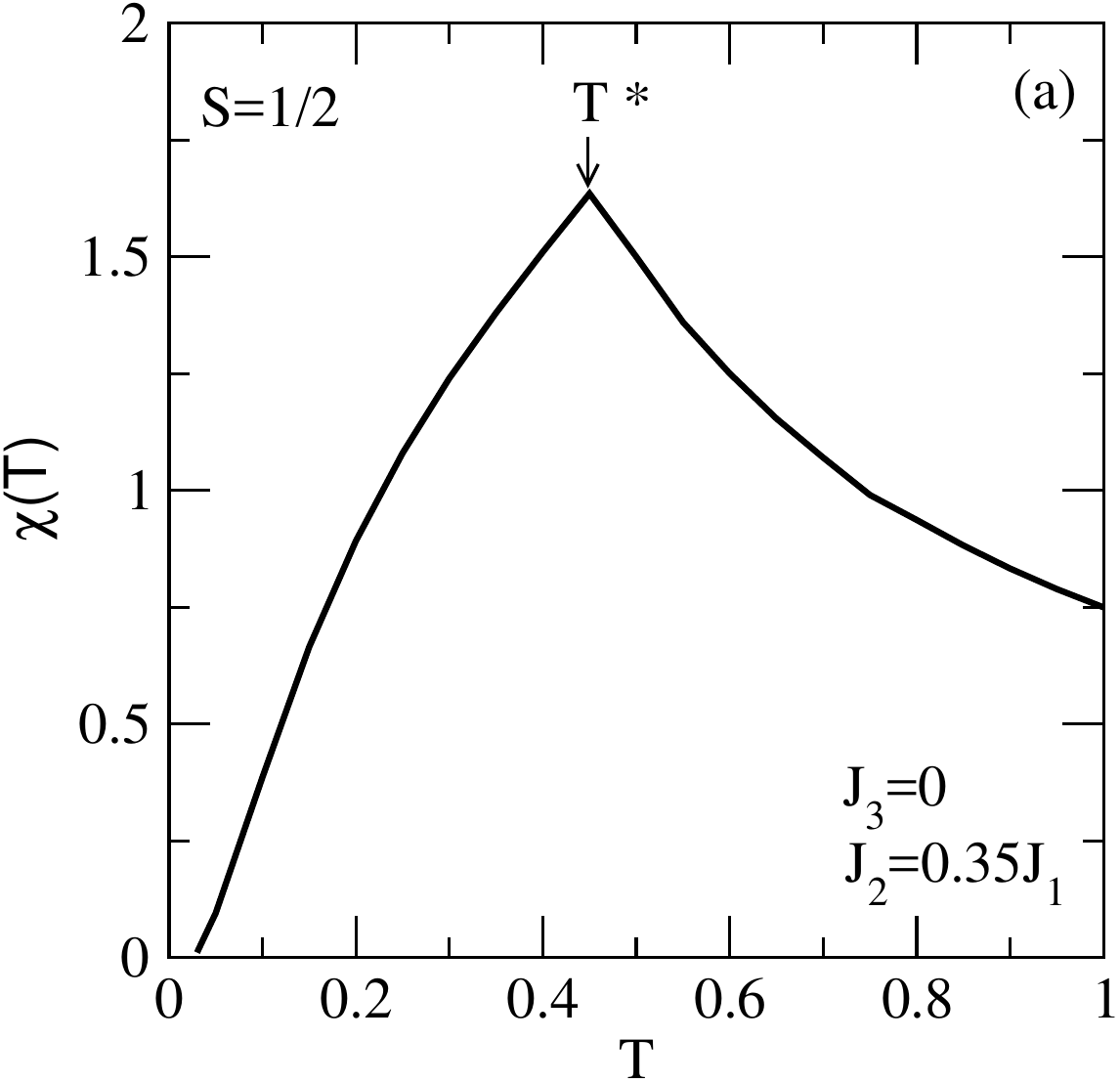}
\includegraphics[width=4cm]{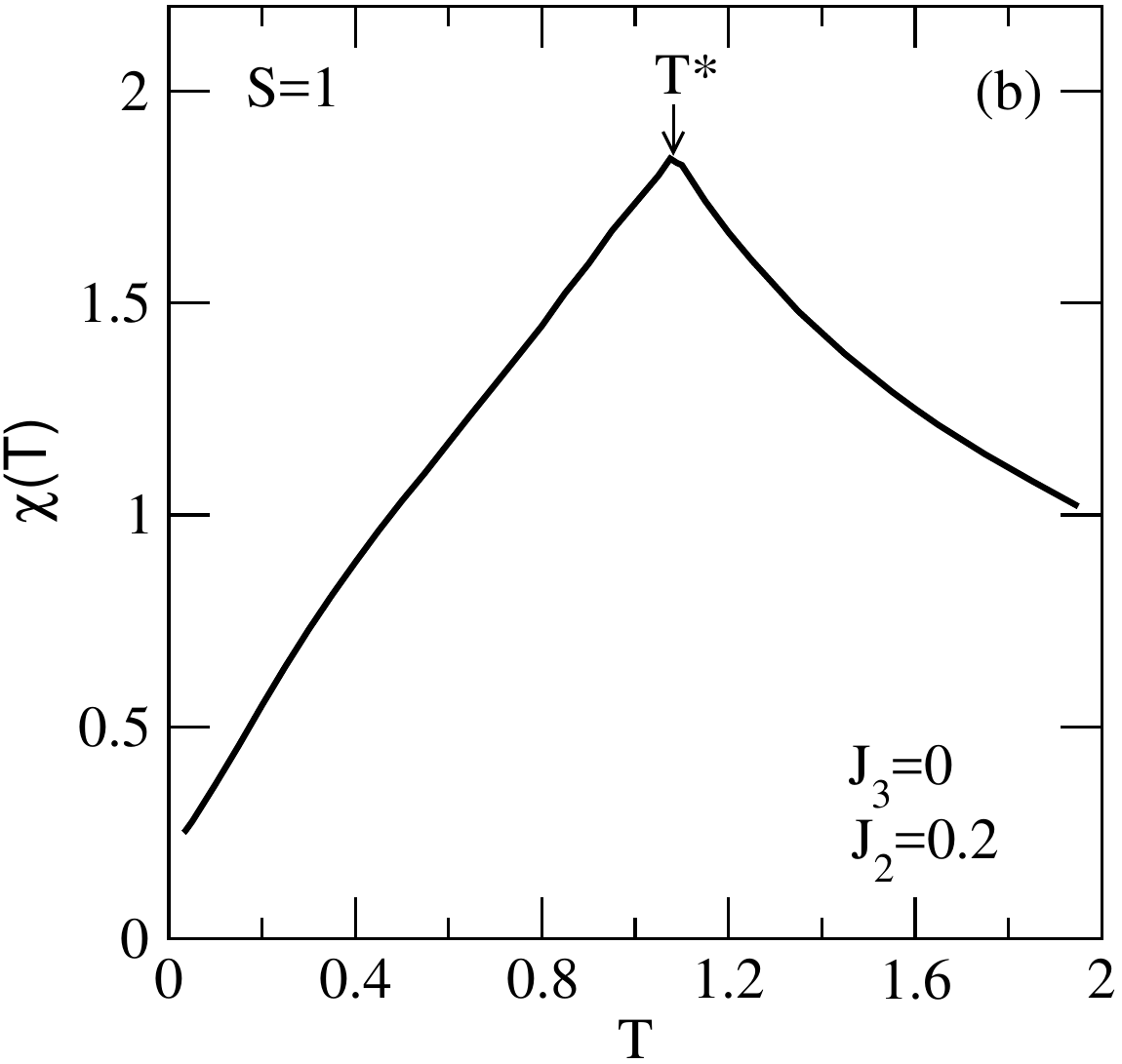}
\includegraphics[width=4cm]{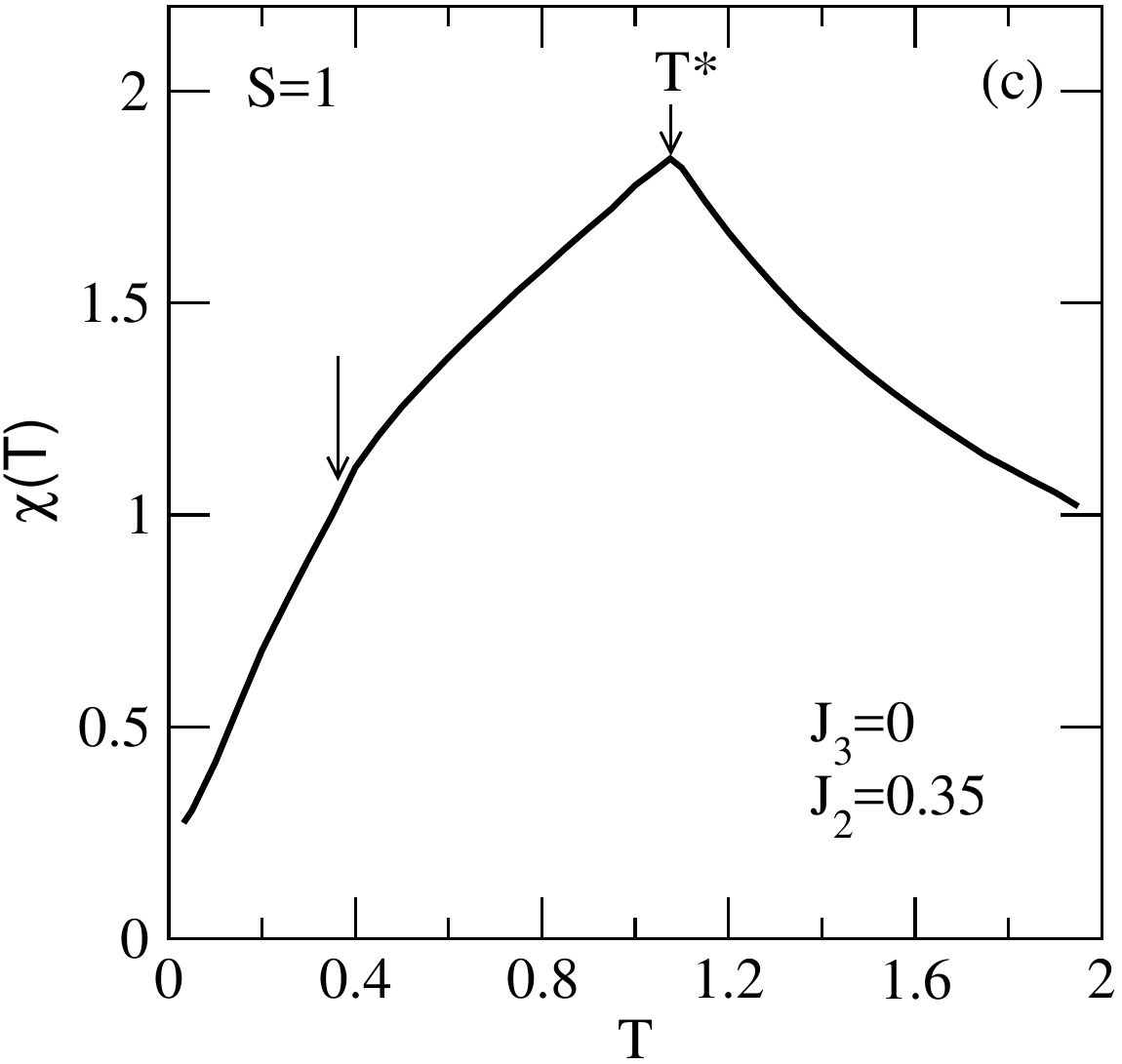}
\includegraphics[width=4cm]{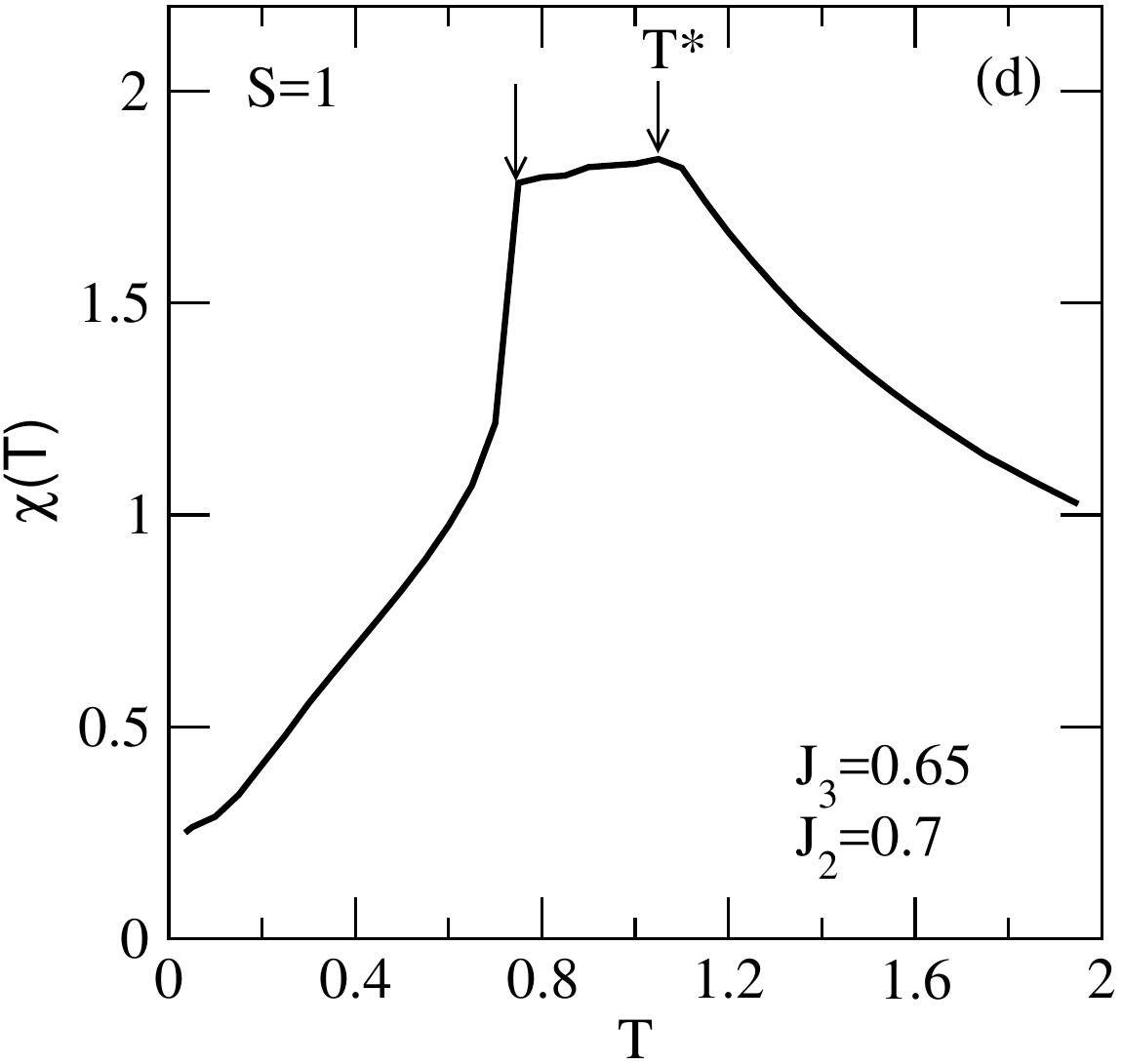}
 \caption{Temperature dependence of magnetic susceptibility for the Heisenberg
model on the honeycomb lattice.  The temperature dependence of $\chi(T)$ is
plotted for the different phases of the model of Fig. \ref{fig:fig0} obtained
at zero temperature $T=0$. In (a) we show the QSL case for $J_3=0$ and
$J_2=0.35 J_1$ for $S=1/2$, in (b) the N\'eel ordered phase for $J_3=0$,
$J_2=0.2 J_1$ for $S=1$, and in (c) the spiral ordered configuration for $J_2=0$,
$J_3=0.35 J_1$ and for the collinear antiferromagnetic phase: $J_2=0.6 J_1$, $J_3=0.7 J_1$.
The vertical arrow in (c) and (d) indicates the temperature at which the 
relative spin orientation changes. We have taken $J_1=1$ in the plot.
} \label{fig:fig1}
\end{figure}

The magnetic susceptibility $\chi(T)$ gives information
about the difficulty of polarizing the spins in the lattice.  We have analyzed
the temperature dependence of the susceptibility using SBMFT, by adding
Bose-Einstein occupation functions in the free energy of Eq.~\ref{eqn:fe} as
well as a weak magnetic field allowing the spin polarization of the Schwinger
bosons. All details are provided in appendices \ref{app:free} and \ref{app:mag}.  We have explored the
finite temperature effects on the different ground states ($T=0$)  of the phase
diagram of Fig. \ref{fig:fig0}. Since we are dealing with a two-dimensional
system with short range interactions, the Mermin-Wagner theorem forbids long
range magnetic order at any finite temperature. Hence, if we raise the
temperature of a $T=0$ magnetically ordered phase, we should expect the opening
of a spin gap and short range correlations. We consider first the case in
which the ground state of the system is the disordered QSL phase in the $S=1/2$
model. As temperature is increased, $\chi(T)$ increases indicating the gradual
destruction of the RVB spin correlations in the QSL. This behavior occurs until
the temperature $T^*$ is reached; at this temperature the system crosses over
to a paramagnet. In the high temperature regime, the magnetic susceptibility
follows a Curie law $\chi(T) \propto 1/T$, as expected for
free (non-interacting) localized spins. This high temperature $T>T^*$ is
encountered in all the parameter ranges explored. If we increase the
temperature from $T=0$ for the N\'eel ordered state, the susceptibility also
raises until $T^*$  is reached due to the gradual reduction of the spatial
extent of the spin correlations with $T$. Above $T^*$ again we find the Curie
behavior as expected. In the case of the spiral phase, we find that the ground
state spiral correlations give way to N\'eel correlations as temperature is
raised above $T \approx  0.4J_1 <T^* \approx 1.1J_1$. A similar behavior is
also found in the CAF phase. At $T \approx 0.75 J_1 $ a transition from short
CAF correlations to N\'eel correlations occur leading to a plateau in the
temperature range $0.75 J_1 < T <  1.1 J_1$ above which the Curie behavior
occurs. These thermally induced changes in the spin orientation indicate the
proximity of the system to a quantum phase transition to another ground state 
with a different magnetic order.

We note that $T^*$ depends strongly on $S$ being enhanced from $T^*=0.45 J_1$
to $T^*=1.1 J_1$ when $S$ is increased from $S=0.5$ to $S=1$ as expected from
the simple mean-field $T^*(S) \propto S(S+1)$ scaling relation.
This leads to $T^*(S=1) =8/3 T^*(S=0.5)$, consistent with our
numerical results. In spite of the similar $T$-dependence of $\chi(T)$ for
$T<T^*$ there are also crucial differences as $T \rightarrow 0$ depending on
whether the ground state of the system is magnetically ordered or not.
When the ground state of the system is either the CAF, spiral
or N\'eel ordered phase, $\chi(T)$, goes to a finite value as $T \rightarrow 0$
as shown in Fig. \ref{fig:fig1} as expected \cite{mila2000}. On the other hand, the $T$-dependence of the QSL is very different
with the susceptibility dropping exponentially to zero\cite{mila2000},
$\chi(T) \propto e^{-\Delta E/k_B T}$, due to the spin-gap $\Delta E$.  Hence, the SBMFT approach is able to describe the whole
$T$-dependence\cite{yu2014} in different ground state configurations of the
spins as shown in Fig. \ref{fig:fig0}.

On the basis of our calculations we discuss some recent magnetic susceptibility
experiments on Na$_2$Co$_2$TeO$_6$, which features a honeycomb lattice of
magnetic  Co$^{2+}$ ions with $S=1/2$.  A magnetic order transition from a
high-temperature Curie paramagnet to a stripe ordered AF (the CAF phase) is
observed\cite{lefrancoise2016} at $T=T_N$. Other features below $T_N$
presumably related to a spin reorientation are observed. In order to capture
long range magnetic order, a three dimensional model consisting of the
$J_1-J_2-J_3$ Heisenberg model describing the honeycomb layers of Co$^{2+}$
atoms coupled through an interlayer, $J_4$ has been considered.  A classical
Monte Carlo evaluation of the model describes the ordering transition at $T_N$
but misses the extra features observed in the magnetic susceptibility at $T<
T_N$.  Our present work shows that the temperature dependence of $\chi(T)$ for
the $J_1-J_2-J_3$ Heisenberg model including quantum fluctuations (within
SBMFT) can be more complex than just the crossover from the Curie paramagnet to
the paramagnet with short correlations, containing a richer structure
associated with changes in the spin orientations induced by temperature.

\section{Dynamic structure factor}
\label{sec:dynam}
			\begin{figure*}[ht]
			\begin{center}
			\includegraphics[width=1\textwidth,clip]{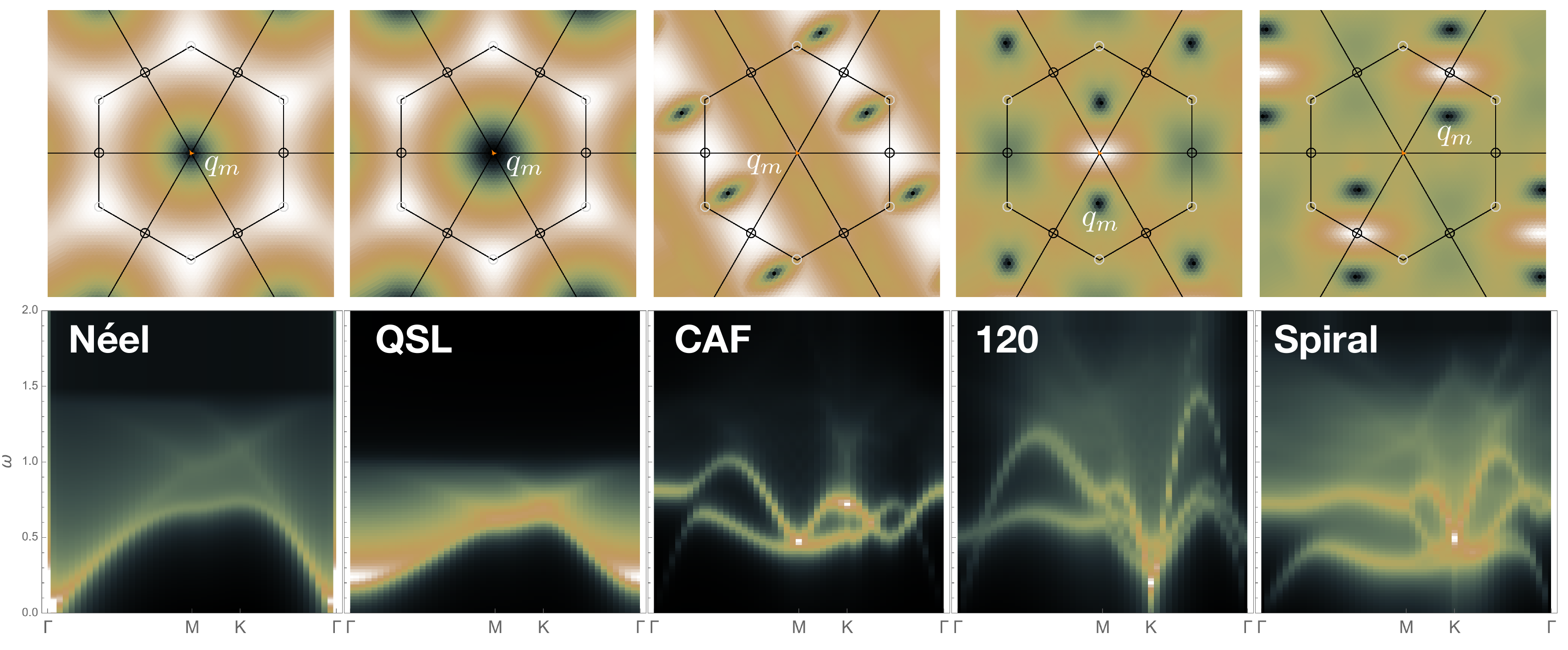}
			\caption{ 
				Dispersion relations and dynamic structure factors along the
path $\Gamma \to M \to K \to \Gamma$ for the five phases of the S=1/2 model
obtained on a $n_c = 48 \times 48$ cluster and discussed in the text. Apart the
quantum spin liquid phase, all phases are ordered with a close gap at ${\bf Q} = 2
{\bf q}_m$ where ${\bf q}_m$ is the minimum of the dispersion relation.
			\label{fig02}
			}
			\end{center}
			\end{figure*}

It is interesting to make connection with neutron experiments and anticipate
what would be the signatures of the various phases we have obtained in our
model. To this purpose, we have also computed the dynamic spin structure factor
defined as
\begin{eqnarray}
	S^{\alpha,\beta}({\bf k},\omega) &=& \frac{1}{n_c} \sum_{i,j} e^{i {\bf k} \cdot ( {\bf r}_i - {\bf r}_j ) } \int_{-\infty}^{+\infty} dt e^{- i \omega t} \langle \vec{S}_i^\alpha(t) \vec{S}_j^\beta \rangle \nonumber \\
\end{eqnarray}
for the $S=1/2$ case only, because of the redundancy of the
phases in the phase diagrams first, but also because we wanted to focus on the
case with the more quantum fluctuations. The expression in terms of the block
elements of the Bogolioubov matrix $T_{\bf q}$ is detailed in Eq.~30 of
[\onlinecite{Halimeh2016}]. In our case, we have derived and diagonalize a
sublattice $2 \times 2$ matrix, the sum of its eigenvalues being plotted in
Fig.~\ref{fig02}, together with the dispersion relation of the lowest band
$\epsilon_{{\bf q}\uparrow}$. We have considered five representative cases, deep in
the domain of each phase, and for a cluster of linear size of 48 unit cells,
large enough to focus on the thermodynamic limit properties. In the $(J_2,J_3)$
plane and for $J_1 = 1$, the N\'eel phase is obtained at $(0.1,0)$, the QSL at
$(0.3,0)$, the CAF at $(0.8,0.8)$, the spiral at $(0.7,0.1)$ and the 120$^o$ at
$(0.8,0.5)$.

Several features appear in these plots.  First, while the signature of the
excitations in the N\'eel phase and the quantum spin liquid (QSL) are quite
similar because no symmetry is broken, the three other phases present clear
distinct features. One has to be careful though, since we plot $S(k,\omega)$
along a specific path connecting high symmetry points of the Brillouin zone,
the incommensurate phases has a Bose condensation of magnons (zero mode in the
energy) at some ${\bf Q}$ vectors that are not necessarily on this
path. As a result, we cannot see the soft modes occurring at these ${\bf Q}$ points,
but only the coherent excitations in the neighboring environment.  This said,
we see that these coherent excitations are quite different for all phases. The
most rigid one, the $120$ triangular phase, has the broadest excitations in
amplitude, while spectral weights on the two others, the CAF and the Spiral
phases, are much smaller than the others.  In the QSL, a small gap is observed
because of the choice of the parameters $(0.3,0.0)$ corresponding to the very
beginning of the gap opening \cite{Lamas}. Also, we see a very high density of
states above the excitation threshold, as expected in a liquid state lacking of
substrate for coherent magnetic excitations.
 
When the gap closes, the system enters the antiferromagnetic N\'eel state, and
one can see a sharpening of the coherent excitations with higher branches. It
is worth noticing that a strong continuous background remains present.  This is
due to quantum fluctuations that lower the net momentum per spin that
one would expect for the classical solution.

\section{Effect of single-ion anisotropy}
\label{sec:singleion}

\begin{figure}[h]
   \centering
    \includegraphics[width=6cm]{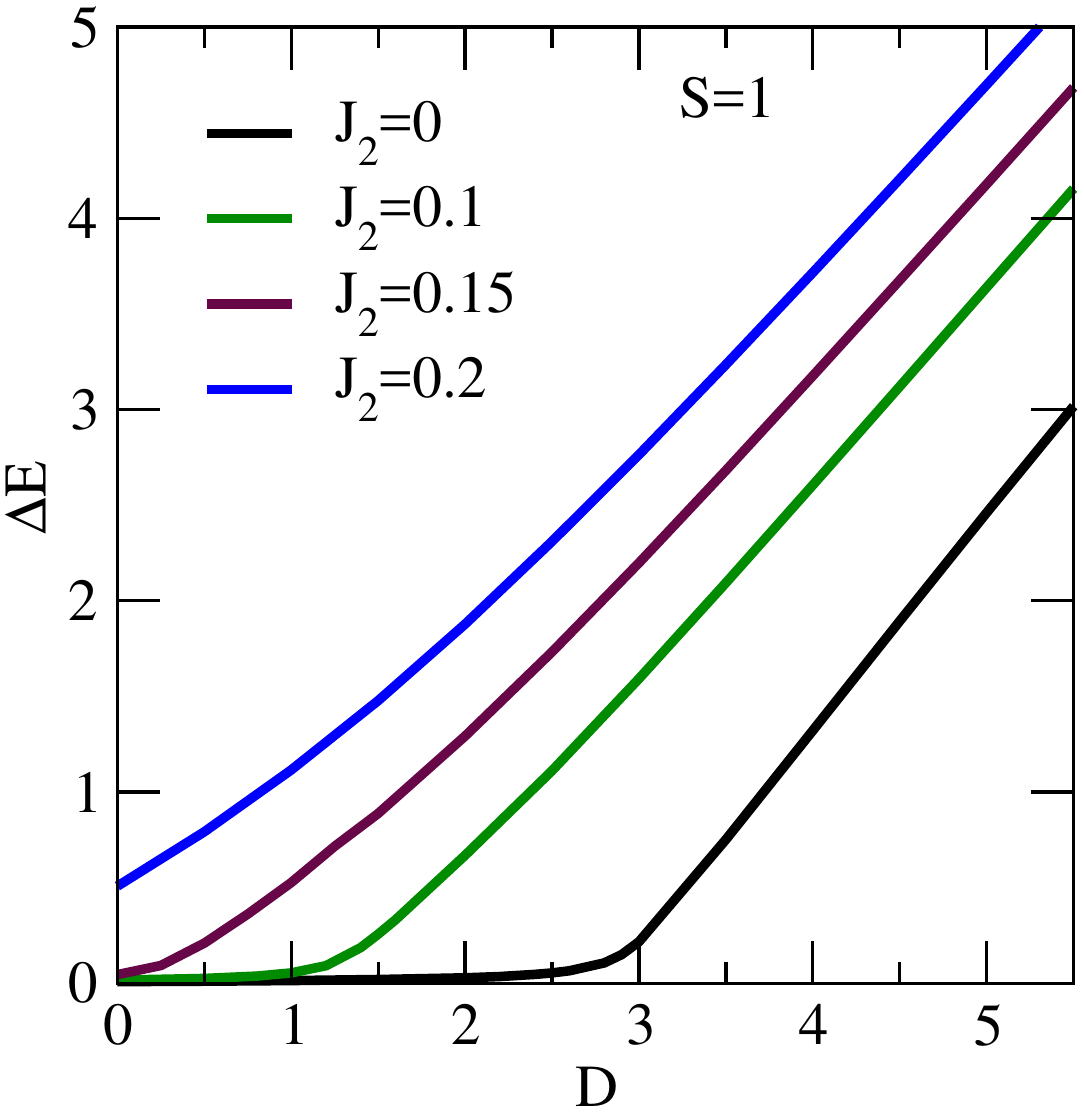}
   \caption{Quantum phase transition from the N\'eel to the large-$D$ phase in the $J_1-J_2$ 
   Heisenberg model on the honeycomb lattice with single-ion anisotropy. The dependence of the gap, $\Delta E$,  with the single-ion anisotropy, $D$, 
    is shown for different frustration strength. A gap opens up and the N\'eel order is suppressed around 
the critical, $D_c$, signalling a quantum phase transition to the large-$D$ phase. The plot shows how $D_c$ is suppressed by the
strength of geometrical frustration, $J_2/J_1$. We have taken $J_1=1$ in the plot. }
 \label{fig:figsu3}
 \end{figure}
 
In real materials, when spins are larger than $1/2$, single-ion anisotropy of
strength $D$ can play a crucial role on the nature of the stabilized phases.
In particular, when $D$ is very large, and positive $D \gg J_{ij}$, a trivial
paramagnetic phase is expected, in competition with the ground state obtained
at zero $D$. It is then important, not only from a theoretical perspective, but
also for making contact with real experiments, to provide the behavior of all
phases upon increasing $D$.  Here, we consider the effect of the single-ion
anisotropy on the N\'eel phase of the honeycomb lattice described by the spin-1
$J_1-J_2$ Heisenberg model on the honeycomb lattice.  $J_3=0$ already includes
the key ingredients between the effect of $D$, magnetic order and magnetic
frustration. 

Hence we consider the $S=1$ model: 
\begin{equation}
H=J_1\sum_{\langle ij\rangle}  {\bf S}_i\cdot {\bf S}_j+ J_2 \sum_{\langle \langle ij\rangle\rangle}  {\bf S}_i\cdot {\bf S}_j + D \sum_i (S_i^z)^2.
\label{eq:heisD}
\end{equation}

In order to properly describe the effect of $D$, the SU(2)
description of the SBMFT is no longer efficient, because the SU(3) symmetries
are not taken into account, neither the {\it magnetic} momenta $0,\pm1$
expected for a spin one. An elegant way of circumventing this problem is to
introduce a SU(3) representation of the spins \cite{Wang2017,Wang2005,pires2015} to deal with this term, as
detailed in Appendix \ref{app:su3}.  Hence, we will apply a SU(3)-SBMFT approach to
analyze the effect of $D$ on the magnetically ordered phases. For simplicity we
consider the above model (in which $J_2,J_3=0$ ) which leads to a N\'eel phase when
$D/J_1=0$.  As said, in the limit of $D/J_1 \gg1$, the ground state is the so-called
large-$D$, a trivial paramagnet which consists on the tensor product of
$S^z_i=0$ on all lattice sites. This can be monitored in our mean field SU(3) approach by having 
a non-zero Bose condensation of bosons carrying the zero magnetic flux.

In Fig. \ref{fig:figsu3} we show the dependence of the gap with $D$ for different $J_2$. 
For $J_2=0$, a  spin gap opens around $D_c \approx  3 J_1 $ signaling the transition from the N\'eel ordered phase to the large-D phase which 
consists on the tensor product of $S^z_i=0$ at all sites. This value is smaller than the $D_c=0.72 (2 z) J_1 \approx 4.3 J_1 $,
where $z=3$ is the coordination of the honeycomb lattice previously obtained in the square lattice. \cite{Wang2005}
As seen from Fig. \ref{fig:figsu3} as $J_2$ is increased, the critical $D_c$ is rapidly suppressed so that the quantum paramagnet phase can be stabilized
at very small $D$, in agreement with previous analytical results. \cite{pires2015}  However, caution is in order
here since the mean-field treatment of the $b^\dagger_{i,0}$ bosons representing the $S^z_i=0$ states at 
each lattice site: $\langle b^\dagger_{i,0} \rangle = \langle  b_{i,0} \rangle =s_0$ provides a reliable description 
of the large-$D$ phase. Hence, we expect a breakdown of the theory when $D \rightarrow 0$. Nevertheless, our
analysis does indicate that a large-$D$ phase can be induced at rather small $D$ in the presence of geometrical 
frustration. 
%obtained through Monte Carlo approaches. 
We conclude that in a $S=1$ honeycomb material with single-ion
anisotropy a quantum phase transition from the
N\'eel ordered state to the quantum paramagnet large-$D$ phase is favored by effectively 
increasing the frustration of the lattice. These results are relevant to the magnetic properties of the layers of Mo$_3$S$_7$(dmit)$_3$ materials which 
realize a $S=1$ honeycomb lattice with single-ion anisotropy, $D$, induced by the spin-orbit 
coupling \cite{merino2016,merino2017}. 

\section{Conclusions}
\label{sec:conclu}

In the present work we provide further evidence for the existence of a QSL in a broad region of the phase diagram 
of the spin-$1/2$, $J_1-J_2-J_3$ Heisenberg model on the honeycomb lattice. This result is consistent with previous
theoretical works which have used numerical approaches. Our result is important since, at present, there is no exact 
method for solving this model and different approximations can, in principle, lead to different results. Also the SBMFT 
is a much more simple and less computationally costly than the heavy numerical approaches already used. 
We also find that when the spin is enlarged to $S=1$, the QSL region disappears and the phase diagram closely resembles the classical
phase diagram indicating the small effect of quantum fluctuations in this case. Hence, our SBMFT analysis suggests that it is
unlikely that a QSL could exist in $S=1$ honeycomb compounds which are described by the $S=1$ $J_1-J_2-J_3$ 
Heisenberg model. 

We have characterized the different ground states  by computing the magnetic susceptibility and dynamic structure factor
which can be directly compared with experimental observations. At large temperatures the Curie behavior
expected for non-interacting localized moments of spin-$S$ is recovered by SBMFT. As temperature is lowered below $T^*$ a suppression
of $\chi(T)$ signaling the onset of short range spin correlations occurs. While for the Z$_2$ spin liquid phase found with 
SBMFT, $\chi(T) \rightarrow 0$, as expected in a spin-gapped state, $\chi(T) \rightarrow const. $ in the magnetically ordered phases. When temperature is increased 
in the CAF phase, we find a jump in the magnetic susceptibility at $T<T^*$ due to a change in the spin orientations induced by temperature. The dynamic structure factor typically 
displays a sharp magnon-like dispersion arising from the triplet combination of two spinons and a weaker background 
associated with the particle-hole type excitations in the spinon continuum. 

The effect of the single-ion anisotropy term is known to be relevant to $S=1$ honeycomb compounds.
For instance, in Ba$_2$NiTeO$_6$ the stripe magnetic ordered structure can be described based on a
$J_1-J_2-J_3$ honeycomb model with $J_3 \lesssim 0$, $J_2/J_1 \sim 2$,  and a relatively large negative $D =-1.4 J_1$ contribution which 
is essential to stabilize the observed stripe order. The presence of $D$ is also crucial to understand the robustness of the stripe phase 
observed when Ni is changed by Co to form Ba$_2$CoTeO$_6$ in spite of the large suppression of the $J_2/J_1=0.5$
ratio estimated from first principles. The effect of single-ion anisotropy is also relevant to the honeycomb layers of the 
organometallic compound, Mo$_3$S$_7$(dmit)$_3$. A transition to the large-$D$ phase can be induced for $D> D_c$,
where $D_c$ is strongly suppressed by the frustration of the lattice. Even if this large $D$-limit cannot be reached it 
would be interesting to analyze the effect of quantum fluctuations on magnetic properties, arising in ordered 
phases close to the quantum disordered large-D phase.

As stated, our work adds further theoretical support in favor of  the existence of a magnetically disordered region in the phase diagram of the $S=1/2$ $J_1-J_2-J_3$
Heisenberg model on the honeycomb lattice. However, the character of such disordered phase is predicted to be different depending on 
the method used. While the SBMFT used here predicts a gapped Z$_2$ quantum spin liquid, ED on small clusters as well as variational Monte Carlo 
approaches\cite{ferrari2017} find a PVBC. Hence, more theoretical work is needed to unambiguously determine the nature of the quantum 
paramagnetic phase. It is highly desirable to extend the phase diagram to finite temperatures for a 
complete comparison with experimental observations and to check the validity of the model for real materials.  

Experimental efforts on searching for a QSL should concentrate on $S=1/2$ honeycomb compounds realizing the $J_1-J_2-J_3$ Heisenberg model 
with $(J_2,J_3)$ in the disordered region of Fig. \ref{fig:fig0}.  Typically most of quasi-two-dimensional honeycomb materials display long range magnetic order
of the N\'eel, spiral and/or CAF type. Exceptions are the $S=3/2$ Bi$_3$Mn$_4$O$_{12}$(NO$_3$) which displays no signs of magnetic order down to 0.4 K,
In$_3$Cu$_2$VO$_9$ or possibly BaCo$_2$(AsO$_4$)$_2$. The $S=3/2-2$ material, CaMn$_2$Sb$_2$, is a N\'eel magnet
which, however, displays coexistent short range magnetic order of different types\cite{mcnally2015}. This unconventional behavior has been 
interpreted in terms of the proximity of this compound to the spiral phase\cite{mazin2013} of the $J_1-J_2-J_3$ classical phase diagram
with $J_3 \approx 0$. Replacing Mn by a lower spin transition metal ion such as Co should enhance quantum fluctuation effects
which could turn the N\'eel state into a QSL state.

\section*{Acknowledgements}
The authors would like to thank S. Fratini and J. Robert for insightful
discussions at the early stage of this work. J.M.  acknowledges financial
support from (MAT2015-66128-R)MINECO/FEDER, UE and A.R. from ANR-ORGANISO
(french national research agency).

\appendix 
% Self-consistent equations and Free energy {{{
\section{Finite temperature Schwinger bosons and Bose condensation}
\label{app:free}

In order to derive the self-consistent equations taking into account Bose
condensates and thermal fluctuations, one has to write down the Free energy
from the diagonalized mean-field Hamiltonian.

The most general mean-field hamiltonian obtained after diagonalization:
\begin{eqnarray*}
H &=& \sum_{{\bf q} , w} \epsilon_{{\bf q}\ua}^{w} \left[ \gamma_{{\bf q}\ua w}^+ \gamma_{{\bf q}\ua w} + \gamma_{-{\bf q}\da w}^+ \gamma_{-{\bf q}\da w} + 1 \right] \\ 
  &+& n_s K( \left\{ {\cal O } \right\}, \left\{ \mu \right\} )
\end{eqnarray*}
with $K( \left\{ {\cal O } \right\}$ a function depending on the mean field parameters and the chemical potentials as in Eq.~\ref{eq:mfh}. The Bogolioubov bosonic
operators $\gamma_{a{\bf q}\ua}^+$ are of dimension 4, 2 for the sublattice $u$ or
$v$, two for the {\it spin} flavor and $n_s$ is the number of unit cells in the
Bravais lattice.

The free energy is defined as:
\begin{eqnarray}
{\cal F} &=& - \frac{1}{\beta} \ln \text{Tr} e^{-\beta H }
\end{eqnarray}
where the trace runs over the number of bosons of type $n_{{\bf q}\ua}^{ w}$ and $n_{-{\bf q}\da}^{w}$. Hence, we have:
\begin{eqnarray}
{\cal F} &=& - \frac{1}{\beta} \ln \text{Tr} e^{-\beta \left[ \sum_{{\bf q} , w} \epsilon^{w}_{{\bf q}\ua} \left( n_{{\bf q}\ua}^{w} + n_{-{\bf q}\da}^{ w} + 1 \right) + n_s K \right] } \nonumber \\
  &=& n_s K + \sum_{{\bf q},w} \epsilon^{w}_{{\bf q}\ua} \nonumber \\ 
  &-& \frac{1}{\beta} \ln \left[\text{Tr} e^{-\beta \sum_{{\bf q} , w} \omega^{w}_{{\bf q}\ua} \left( n_{{\bf q}\ua}^{w} + n_{-{\bf q}\da}^{w} \right) }\right] \nonumber \\
  &=& n_s K + \sum_{{\bf q},w} \epsilon^{w}_{{\bf q}\ua} + \frac{2}{\beta} \sum_{{\bf q},w} \ln \left[  1 -  e^{-\beta \epsilon^{w}_{{\bf q}\ua} }\right]
\end{eqnarray}

From the free energy per unit-cell $f = {\cal F} / n_s$, one can derive the self-consistent (SC)
equations at finite temperature and dependent on the Bose-Einstein occupation function
$n_{\text{BE}}(\epsilon_{{\bf q}\ua})$:

\begin{eqnarray}
\frac{\partial f}{\partial \alpha } &=& \frac{\partial K}{\partial \alpha } + \frac{1}{n_s} \sum_{{\bf q},w} \frac{\partial \epsilon^{w}_{{\bf q}\ua} }{\partial \alpha } +  \frac{2}{n_s \beta} \sum_{{\bf q},w} \frac{\partial \ln \left[  1 -  e^{-\beta \epsilon^{w}_{{\bf q}\ua} }\right] }{\partial \alpha } = 0 \nonumber \\
\end{eqnarray}
where $\alpha$ is one of the mean-field parameter $\left\{ {\cal O}  \right\}$ and $\left\{ {\mu}  \right\}$.
By noticing that:
\begin{eqnarray}
\frac{\partial \ln \left[  1 -  e^{-\beta \epsilon^{w}_{{\bf q}\ua} }\right] }{\partial \alpha } &=& \beta \frac{\partial \epsilon^{w}_{{\bf q}\ua}}{\partial \alpha} n_{\text{BE}}( \epsilon^{w}_{{\bf q}\ua} ),
\end{eqnarray}
one obtain the following SC equations:
\begin{eqnarray}
- \frac{\partial K}{\partial \alpha } &=& \frac{1}{n_s} \sum_{{\bf q},w} \frac{\partial \epsilon^{w}_{{\bf q}\ua}}{\partial \alpha } \left[ 1 + 2 n_{\text{BE}}( \epsilon^{w}_{{\bf q}\ua} ) \right] 
\label{eq:sc}
\end{eqnarray}

Due to Mermin-Wagner, there is no phase transition in 2D systems at finite
temperature, thus the dispersion relation is always gapped and the ground state
is disordered. At $T=0$ however, bosons can condense and one has to properly
take into account Bose condensates in the SC equations.  Bose condensates
appear as soon as the dispersion relation presents soft modes
$\epsilon^{w}_{\xi_n^{w}} = 0$ for any of the $n_c^w$ {\bf q} points $\xi_n^{w}$. We
have explicitely displayed the sublattice index because in the case of several
bands, only certains can have a zero mode energy.

Note that it is also possible to extract from these equations the $T=0$
expression of the SC equations taking into account the presence of Bose
condensates. Such condensates usually relate to the fact that a symmetry
breaking is obtained precisely at these $\xi_n^{w}$. Thus, it is possible to
obtain the $T=0$ limit simply by imposing the following condition extracted
from the boson density constraint on a sublattice $w$:
\begin{eqnarray}
\nu^w_B &=& \frac{1}{n_s} \sum_{{\bf q}} \frac{\partial \epsilon^{w}_{{\bf
q}\ua}}{\partial \mu   } n_{\text{BE}}( \epsilon^{w}_{{\bf q}\ua} ) \nonumber
\\ &\Rightarrow& \frac{1}{n_s}\frac{\partial \epsilon^{w}_{{\bf
q}\ua}}{\partial \mu   } n_{\text{BE}}( \epsilon^{w}_{{\bf q}\ua} ) \to \frac{
\nu^w_B }{n_c^w} \sum_{n=1}^{n_c^w} \delta( k - \xi_n^{w})
\end{eqnarray}
and thus:
\begin{eqnarray}
n_{\text{BE}}( \epsilon^{w}_{{\bf q}\ua} ) \to \nu^w_B\frac{n_s}{ n_c^w} \frac{
1  }{ \frac{\partial \epsilon^{w}_{{\bf q}\ua}}{\partial \mu   } }
\sum_{n=1}^{n_c^w} \delta( {\bf q} - \xi_n^{w}).
\end{eqnarray}

Plugging this expression in the general form of the SC equation, we obtain finally:
\begin{eqnarray}
-\frac{\partial K }{\partial \alpha} &=& \frac{1}{n_s} \sum_{{\bf q},w}
\frac{\partial \epsilon^{w}_{{\bf q}\ua}}{\partial \alpha   } + 2 \sum_{w}
\frac{ \nu^w_B }{n_c^w} \sum_{n=1}^{n_c^w} \frac{ \frac{\partial
\epsilon^{w}_{\xi_n^{w} }}{\partial \alpha   }  }{ \frac{\partial
\epsilon^{w}_{\xi_{n_s}}}{\partial \mu   } }.
\end{eqnarray}

Here, one has to be careful. Indeed, at each $\xi$ point, there are two branches in
the dispersion relation that are ordered from the lowest to the highest eigenenergy.
Usually, the lowest branch reaches the zero but not the highest, and thus only one type
of exists. However, it is not excluded that the two branches
reach zero at the same $\xi$ points, hence both condensates should
exist. Note that it is unlikely, but it has to be checked, that the two
types of condensates appear at different $\xi$ points. Finally, in the case
for which only one type of the condensate appears, only this type has to be
considered in the previous equation.
% }}}

\section{Effect of a magnetic field}
\label{app:mag}

Here we show the extension of the finite temperature SU(2) SBMFT formalism to include a magnetic field. 
The equations described below are used to compute the $T$-dependence of the magnetic susceptibility, $\chi(T)$, 
discussed in the paper. 

The SBMFT under a uniform magnetic field, $B$, along the $z$-axis reads:
\begin{equation}
H(B)=H-{\mu_B B \over 2} \sum_i \left( \langle b^\dagger_{i\uparrow} b_{i\uparrow} \rangle - \langle b^\dagger_{i\downarrow}b_{i\downarrow}\rangle \right), 
\end{equation}
where $H$ is the SBMFT hamiltonian without the magnetic field introduced in Eq. (\ref{eq:mfh}).   
Since the magnetic field just leads to a different chemical potential for the $\uparrow$ and 
$\downarrow$ bosons: $\mu_\sigma=\mu-2\sigma \mu_b B/2 $, where $\sigma=\pm {1 \over 2}$, we can 
just replace $\mu$ by $\mu_\sigma$ leading to different spinon dispersions: $\epsilon_{{\bf q}\ua}^{w} \neq \epsilon_{{\bf q}\da}^w$
when $B \neq 0$. 
Hence, the diagonalized hamiltonian in the presence of the magnetic field,  $B$, now reads:
\begin{eqnarray*}  
H(B) &=& \sum_{{\bf q} , \sigma, w} \epsilon_{{\bf q}\sigma}^{w} \left[ \gamma_{{\bf q}\sigma w}^+ \gamma_{{\bf q}\sigma w} +{1 \over 2} \right]  + n_s K( \left\{ {\cal O } \right\}, \left\{ \mu \right\} ).
\end{eqnarray*}
Following the same analysis as in the previous section but using $H(B)$ instead of $H$ we arrive 
at the following SC equations:
\begin{eqnarray}
- \frac{\partial K}{\partial \alpha } &=& \frac{1}{n_s} \sum_{{\bf q},w} \frac{\partial\sigma \epsilon^{w}_{{\bf q}\sigma}}{\partial \alpha } \left[ {1 \over 2} + n_{\text{BE}}( \epsilon^{w}_{{\bf q}\sigma} ) \right],
\end{eqnarray}
which recovers the SC equations (\ref{eq:sc}) when $B=0$.
By evaluating the uniform magnetization, $m_B(T)$, induced by a small field, $B$, and taking $B \rightarrow 0$ we can obtain the 
temperature dependence of the magnetic susceptibility:
\begin{equation}
 \chi(T)=\lim_{B \rightarrow 0} {\partial m_B(T)  \over \partial B},
 \end{equation}
where the uniform magnetization reads:
\begin{equation}
m_B(T)=\sum_{{\bf q},w} \sigma n_{\text{BE}}( \epsilon^{w}_{{\bf q}\sigma} ).
\end{equation}

\section{SU(3) formulation of the Heisenberg model}
\label{app:su3}

Following [\onlinecite{Wang2017,Wang2005}], in the SU(3) formulation of the $J_1-J_2$ Heisenberg model (\ref{eq:heisD}) we introduce three Schwinger bosons:
\begin{eqnarray}
|1 \rangle &=& b^\dagger_{i,+} |0\rangle
\nonumber \\
|0 \rangle &=&b^\dagger_{i,0} |0 \rangle
\nonumber \\
|-1 \rangle &=&b^\dagger_{i,-} |0 \rangle,
\end{eqnarray} 
whIch represent the $S^z=-1,0,+1$ projection of the $S=1$ at each site. The Schwinger bosons 
at each site satisfy the constraint: 
\begin{equation}
b^\dagger_{i,+} b_{i,+} + b^\dagger_{i,0} b_{i,0} + b^\dagger_{i,-} b_{i,-}=1.
\end{equation}
The spin operators can be expressed in terms of the Schwinger bosons as:
\begin{eqnarray}
S^+_i&=&\sqrt{2} (b^\dagger_{i,0} b_{i,-} + b^\dagger_{i,+} b_{i,0})
\nonumber \\
S^-_i&=&\sqrt{2} (b^\dagger_{i,-} b_{i,0}+ b^\dagger_{i,0} b_{i,+})
\nonumber \\
S^z_i&=&b^\dagger_{i,+} b_{i,+} - b^\dagger_{i,-} b_{i,-}.
\end{eqnarray} 
Introducing these operators and assuming the condensation of the 0-bosons:
$\langle b^\dagger_{i,0} \rangle = \langle  b_{i,0} \rangle =s_0$, 
the hamiltonian reads:
\begin{eqnarray}
H &=& J_1 s_0^2\sum_{\langle ij \rangle} \{ (b^\dagger_{i,-1} b_{j,-1} + b_{i,-1} b^\dagger_{j,-1})\nonumber \\ &+&  (b^\dagger_{i,+1} b_{j,+1} + b_{i,+1} b^\dagger_{j,+1})
\nonumber \\
&+&(b^\dagger_{i,+1} b^\dagger_{j,-1} + b^\dagger_{i,-1} b^\dagger_{j,+1})
\nonumber \\ &+&  (b_{i,+1} b_{j,-1} + b_{i,-1} b_{j,+1}) \} 
\nonumber \\
&+&J_1 \sum_{\langle ij \rangle} (b^\dagger_{i,+1} b_{i,+1} - b^\dagger_{i,-1} b_{i,-1} ) \nonumber \\ &\times& (b^\dagger_{j,+1} b_{j,+1} - b^\dagger_{j,-1} b_{j,-1})
\nonumber \\
&+&J_2 s_0^2\sum_{\langle\langle ij \rangle\rangle} \{ (b^\dagger_{i,-1} b_{j,-1} + b_{i,-1} b^\dagger_{j,-1})
\nonumber \\ &+&  (b^\dagger_{i,+1} b_{j,+1} + b_{i,+1} b^\dagger_{j,+1})
\nonumber \\
&+&(b^\dagger_{i,+1} b^\dagger_{j,-1} + b^\dagger_{i,-1} b^\dagger_{j,+1})
\nonumber \\ &+&  (b_{i,+1} b_{j,-1} + b_{i,-1} b_{j,+1}) \} 
\nonumber \\
&+&J_2 \sum_{\langle \langle ij \rangle \rangle} (b^\dagger_{i,+1} b_{i,+1} - b^\dagger_{i,-1} b_{i,-1} ) \nonumber \\ &\times&(b^\dagger_{j,+1} b_{j,+1} - b^\dagger_{j,-1} b_{j,-1})
\nonumber \\
&+& D \sum_i (b^\dagger_{i,+1} b_{j,+1} + b^\dagger_{i,-1} b_{i,-1})^2.
\nonumber \\
\end{eqnarray} 
We can treat the remaining quartic terms by using a further mean-field decoupling so that:
\begin{eqnarray}
(b^\dagger_{i,+1} b_{i,+1} - b^\dagger_{i,-1} b_{i,-1} ) (b^\dagger_{j,+1} b_{j,+1} - b^\dagger_{j,-1} b_{j,-1}) &=&
\nonumber \\
{1 \over 2 } (1-s_0^2) (b^\dagger_{i,+1} b_{i,+1} + b^\dagger_{j,+1} b_{j,+1} )
\nonumber \\
+ {1 \over 2} (1-s_0^2)(b^\dagger_{i,-1} b_{i,-1} + b^\dagger_ {j,-1} b_{j,-1} )
\nonumber \\
- p_\delta (b_{i,+1} b_{j,-1} + b^\dagger_{i,+1} b^\dagger_{j,-1} + b_{i,-1} b_{j,+1} + b^\dagger_{i,-1} b^\dagger_{j,+1} ) 
\nonumber \\
-{1 \over 2} (1-s_0^2)^2 + 2 p_\delta^2, \nonumber \\
\end{eqnarray}
where the real mean-field parameter is: $p_\delta=\langle b^\dagger_{i,-1}  b^\dagger_{j,+1}  \rangle= \langle b_{i,-1} b_{j,+1} \rangle$,
with $\delta=1,2$ for the nearest and next-nearest neighbors, respectively. 

Fourier transforming the bosons: $b_{{\bf r},w}={1 \over \sqrt{n_c}} \sum_{\bf q} e^{i {\bf q} {\bf r} } b_{{\bf q}, w}$, where $n_c$ is the number of 
cells ($n_s=2 n_c$) the final mean-field Hamiltonian reads:
\begin{eqnarray}
H=\sum_{\bf q} \Psi^\dagger_{\bf q} M_{\bf q} \Psi_{\bf q}-\sum_{\bf q} (M^{33}_{\bf q} +M^{44}_{\bf q} )+C, \nonumber \\
\end{eqnarray} 
where:
\begin{eqnarray}
M_{\bf q} &=&
  \begin{bmatrix}
	A_{\bf q } & B_{\bf q } \\
	B_{\bf q } & A_{\bf q } 
  \end{bmatrix}, \nonumber \\
A_{\bf q} &=&
  \begin{bmatrix}
   \tilde{\mu}  & J_1 s_0^2 \gamma_{1{\bf q}} \\
    J_1 s_0^2 \gamma^*_{1{\bf q}}  & \tilde{\mu} 
     \end{bmatrix}, \nonumber \\ B_{\bf q} &=&
  \begin{bmatrix}
   J_2 (s_0^2 -p_2)  \gamma_{2 {\bf q}}& J_1 (s_0^2-p_1) \gamma_{1{\bf q}} \\
   J_1 (s_0^2 -p_1)  \gamma^*_{1 {\bf q}} &J_2 (s_0^2 -p_2)  \gamma_{2 {\bf q}} 
     \end{bmatrix},
\end{eqnarray}
and
$
\tilde{\mu}=\mu +{3 \over 2} (1- s_0^2) (J_1+2 J_2)+2 J_2 s_0^2 \gamma_{2{\bf q}}+D.
$
The dispersion relations read: 
\begin{eqnarray}
\gamma_{1{\bf q}} &=& 1 + e^{i (k_1-k_2)} + e^{-i k_2},\\
\gamma_{2{\bf q}} &=& \cos(k_1)+\cos(k_2)+\cos(k_1-k_2),
\end{eqnarray}
and finally:
\begin{eqnarray}
\frac{C}{n_c} &=& 3 J_1 \left[2p_1^2 - {1 \over 2}(1-s_0^2)^2\right] \nonumber \\ &+& 6 J_2 \left[2p_2^2 - {1 \over 2}(1-s_0^2)^2\right] +2 \mu(s_0^2-1).
\end{eqnarray}

Diagonalization of $\sigma_z \cdot M_{\bf q}$ leads to the Bogoliubov quasiparticle
dispersions as described in the main text. The ground state
energy per site can then be expressed in terms of these new Bogoliubov
quasiparticles as: 
\begin{equation}
e_0={E_0 \over 2 n_c}=
\nonumber\\
{1 \over 2 n_c} \left[ \sum_{{\bf q},\omega} \epsilon^\omega_{{\bf q}\ua} - \sum_{\bf q} (M^{33}_{\bf q}+M^{44}_{\bf q} )+C \right],
\end{equation}
where $\omega=1,2$ denotes the two quasiparticle dispersions.
The SC equations obtained from the minimization of the total energy are:
\begin{eqnarray}
p_1 &=& -{1 \over 12 J_1 n_c} \sum_{{\bf q},\omega} { \partial \epsilon^\omega_{{\bf q} \ua}  \over \partial p_1  }
\nonumber \\
p_2 &=& -{1 \over 24 J_2 n_c} \sum_{{\bf q},\omega} { \partial \epsilon^\omega_{{\bf q} \ua}  \over \partial p_2  }
\nonumber \\
2-s_0^2&=& {1 \over 2 n_c} \sum_{{\bf q},\omega} { \partial \epsilon^\omega_{{\bf q} \ua}  \over \partial \mu  }
\nonumber \\
\mu &=& -2 J_1-{7 \over 2} J_2+ {3 s_0^2 \over 2 } J_1 +3 s_0^2 J_2- {1 \over  2 n_c} \sum_{{\bf q},\omega} {\partial \epsilon^\omega_{{\bf q}\ua}  \over \partial s_0^2  }.
\nonumber\\
\end{eqnarray}


\begin{thebibliography}{50}%
\bibitem{balents} L. Balents, Nature {\bf 464} 199 (2010).
\bibitem{Anderson1973} P. .W. Anderson, Mat. Res. Bull {\bf 8}, 153 (1973).
\bibitem{Fazekas1974} P. Fazekas and P. .W. Anderson, Phil. Mag. {\bf 30}, 423 (1974).
\bibitem{Anderson1987} P. .W. Anderson, Science {\bf 235}, 1196 (1987).
\bibitem{Liang1988} S. Liang, B. Dou\c{c}ot and P. .W. Anderson, \prl {\bf 61}, 365 (1988).
\bibitem{Sachdev1992} S. Sachdev, \prb {\bf 45}, 12377  (1992).
\bibitem{Sandvik2005} A. W. Sandvik, \prl {\bf 95}, 207203 (2005).
\bibitem{merino1999} J. Merino, R. H. Mckenzie, J. B. Marston, and C. H. Chung, J. Phys.: Condens. Matter {\bf 11} 2965 (1999).
\bibitem{trumper1999} A. E. Trumper, Phys. Rev. B {\bf  60}, 2987 (1999).
\bibitem{merino2014} J. Merino, M. Holt, and B. J. Powell, Phys. Rev. B {\bf 89}, 245112 (2014).
\bibitem{holt2014} M. Holt, B. J. Powell, and R. H. McKenzie, Phys. Rev. B {\bf 89}, 174415 (2014).
\bibitem{kanoda2016} Y. Zhou, K. Kanoda, T.-K Ng, Rev. Mod. Phys. {\bf 89} 025003 (2017).
\bibitem{norman2017} M. Norman, Rev. Mod. Phys. {\bf 88}, 041002 (2016). 
\bibitem{lefrancoise2016} E. Lefrancoise, et al. Phys. Rev. B {\bf 94}, 214416 (20016).
\bibitem{martin2012} N. Martin, L.-P Regnault, and S. Klimko, Jour. Phys. Conf. Ser. {\bf 340} 012012 (2012).
\bibitem{smirnova2009} O. Smirnova {\it et. al.} J. Am. Chem. Soc. {\bf 131}, 8313 (2009).
\bibitem{yan2012} Y. J. Yan {\it et. al.}, Phys. Rev. B {\bf 85} 85102 (2012).
\bibitem{banerjee2016}  A. Banerjee, {\it et. al.} Nat. Mater. {\bf 15}, 733 (2016).
\bibitem{banerjee2017} A. Banerjee, {\it et. al.}, Science {\bf 356}, 1055 (2017).
\bibitem{do2017} S.-H. Do, {\it et. al.}, Nat. Phys. {\bf 13} 1079 (2017).
\bibitem{kitaev2005} A. Kitaev, Ann. of Phys. {\bf 321} 2 (2005).\bibitem{asai2016} S. Asai,M. Soda, K. Kasatani, T. Ono, M. Avdeev, and T.Masuda, Phys. Rev. B {\bf 93}, 024412 (2016).
\bibitem{asai2017} S. Asai, {\it et. al.},  arXiv:1708.08717v1.
\bibitem{merino2016} J. Merino, A. C. Jacko, A. L. Khosla, and B. J. Powell, Phys. Rev. B {\bf 94}, 205109 (2016). 
\bibitem{khosla2017} A. L. Khosla, A. C. Jacko, J. Merino, and B. J. Powell,  Phys. Rev. B {\bf 95}, 115109 (2017).
\bibitem{merino2017} J. Merino, A. C. Jacko,  A. L. Khosla, and B. J. Powell, Phys. Rev. B {\bf 96}, 205118 (2017).
\bibitem{jacko2017} A. C. Jacko, A. L. Khosla, J. Merino, and B. J. Powell,  Phys. Rev. B {\bf 95}, 155120 (2017).
\bibitem{meng2010} Z. Y. Meng, T. C. Lang, S. Wessel, F. F. Assaad, A. Muramatsu, Nature {\bf 464}, 847 (2010).
\bibitem{sorella2012} S. Sorella, Y. Otsuka, S. Yunoki, Scientific Reports  {\bf 2}, 992 (2012).
\bibitem{lhuillier2001} J.B. Fouet, P. Sindzingre, and C. Lhuillier, Eur. Phys. J B {\bf 20}, 241 (2001).
\bibitem{albuquerque2011} A. F. Albuquerque, D. Schwandt, B. Het\'enyi, S. Capponi,M. Mambrini, and A. M. L\"auchli, Phys. Rev. B {\bf 84}, 024406 (2011).
\bibitem{reuther2011} J. Reuther, D. A. Abanin, and R. Thomale, Phys. Rev. B {\bf 84}, 014417 (2011).
\bibitem{sheng2013} S.-S. Gong, D. N. Sheng, O. I. Motrunich, and M. P. A. Fisher, Phys. Rev. B {\bf 88}, 165138 (2013).
\bibitem{oitmaa2011} J. Oitmaa and R. R. P. Singh, Phys. Rev. B {\bf 84}, 094424 (2011).
\bibitem{auerbach} A. Auerbach, {Interacting Electrons and Quantum Magnetism}, Springer-Verlag (1994).
\bibitem{Lamas} H. Zhang and C. A. Lamas, Phys. Rev. B {\bf 87}, 024415 (2013).
\bibitem{sheng2015} S.-S. Gong, W. Zhu, and D. N. Sheng, Phys. Rev. B {\bf 92}, 195110 (2015).
\bibitem{gazza1993} C. J. Gazza and H. A. Ceccatto, J. Phys: Condens. Matter {\bf 5}, L135 (1993).
\bibitem{flint2009} R. Flint and P. Coleman, Phys. Rev. B {\bf 79}, 014424 (2009).
\bibitem{jaime2014} V. Kapf, M. Jaime, and C. D. Batista, Rev. Mod. Phys. {\bf 86}, 563 (2014). 
\bibitem{lamas2013} H. Zhang, and C. A. Lamas, Phys. Rev. B {\bf 87}, 024415 (2013).
\bibitem{cabra2011} D. C. Cabra, C. A. Lamas, and H. D. Rosales, Phys. Rev. B {\bf 83}, 094506 (2011).
\bibitem{Wang2010} Fa Wang, \prb {\bf 82}, 024419 (2010).
\bibitem{Halimeh2016} J. C. Halimeh and M. Punk,  \prb {\bf 94}, 104413 (2016).
\bibitem{Colpa} J. H. P. Colpa, Phys. A {\bf 93}, 327 (1978).
\bibitem{Toth} S. Toth and B. Lake, Journal of Physics: Condens. Matter {\bf 27}, 166002 (2015).
\bibitem{rastelli1979} E. Rastelli, A. Tassi, and L. Reatto, Physica {\bf 97B}, 1 (1979).
\bibitem{mulder2010} A. Mulder, R. Ganesh, L. Capriotti, and A. Paramekanti, Phys. Rev. B {\bf 81} , 214419 (2010).
\bibitem{white2013} Z. Zhu, D. A. Huse, and S. R. White, Phys. Rev. Lett. {\bf 110}, 127205 (2013).
\bibitem{bishop2012} P. H. Y. Li, R. F. Bishop, D. J. J. Farnell, and C. E. Campbell, Phys. Rev. B {\bf 86}, 144404 (2012).
\bibitem{bishop2016} P. H. Y. Li and R. F. Bishop, Phys. Rev.  B {\bf  93}, 214438 (2016).
\bibitem{mila2000} F. Mila, Eur. J. Phys. {\bf 21} 499 (2000).
\bibitem{yu2014} X.-L Yu, D.-Yong Liu, P. Li, L.-J Zou, Phys. E {\bf 59}, 41 (2014).
\bibitem{Wang2017} Zhentao Wang, Adrian E. Feiguin, Wei Zhu, Oleg A. Starykh, Andrey V. Chubukov, and Cristian D. Batista, \prb {\bf 96}, 184409 (2017).
\bibitem{Wang2005} H.-T. Wang and Y. Wang, Phys. Rev B {\bf 71}, 104429 (2005).
\bibitem{pires2015} A. S. T. Pires, Phys. B, {\bf 479}, 130 (2015).
\bibitem{ferrari2017} F. Ferrari, S. Bieri, and F. Becca, Phys. Rev. B {\bf 96}, 104401 (2017).
\bibitem{mcnally2015} D. E. McNally, {\it et. al.}, Phys. Rev. B {\bf 91},  180407 (R) (2015).
\bibitem{mazin2013}  I. Mazin, arXiv:1309.3744.
%\bibitem{powell2016} B. J. Powell, J. Merino, A. L. Khosla, A. C. Jacko, Phys. Rev. B {\bf 95}, 094432 (2017); Phys. Rev. B \textbf{96}, 099902 (2017).
%\bibitem{savary} L. Savary and L. Balents, Rep. Prog. Phys. {\bf 80} 016502 (2017).
%\bibitem{mezio2012} A. Mezio, L. O. Manuel, R. R. P. Singh, and A. E. Trumper, New J. Phys. {\bf 14}, 123033 (2012).
\end{thebibliography}
\end{document}